\documentclass[11pt]{article}
\usepackage[T1]{fontenc}
\usepackage{txfonts}
\usepackage{graphicx}
\usepackage{natbib}
\usepackage{afterpage}

\newtheorem{prop}{Proposition}
\newtheorem{rem}[prop]{Remark}
\newtheorem{lem}[prop]{Lemma}
\newtheorem{exmp}[prop]{Example}

\newcommand{\set}[1]{\{ #1 \}}

\newcommand{\proofstart}{\noindent\textbf{Proof.\ }}
\newcommand{\qedmark}{\rule{1ex}{1ex}}
\newcommand{\proofend}{\hspace*{\fill}\qedmark\ }

\newcommand{\ev}{\mathrm{ev}}
\newcommand{\wt}{\mathrm{wt}}
\newcommand{\RR}{\mathcal{L}(\infty Q)}

\begin{document}
\hypersetup{pdfstartview={FitBH -32768},pdfauthor={Ryutaroh Matsumoto, Diego Ruano and Olav Geil},pdftitle={List Decoding Algorithm based on Voting in Gr\"obner Bases for General One-Point AG Codes},pdfkeywords={algebraic geometry code, Gr\"obner basis, list decoding}}
\author{Ryutaroh Matsumoto\thanks{%
Department of Communications and Computer Enginnering,
Tokyo Instiutte of Technology, 152-8550 Japan.},\ Diego Ruano\thanks{%
Department of Mathematical Sciences, Aalborg University, Denmark.},\ 
and 
Olav Geil\footnotemark[2]}
\title{List Decoding Algorithm based on Voting in Gr\"obner Bases
  for General One-Point AG Codes\thanks{Accepted for publication
    in Journal of Symbolic Computation. The proposed
    algorithm in this paper was published
without any proof of
its correctness in Proc.\  2012 IEEE International Symposium on Information Theory,
Cambridge, MA, USA, July 2012, pp.\ 86--90 \citep{gmr12isit}.}}
\date{February, 2016}
\maketitle
\begin{abstract}
We generalize the
unique decoding algorithm for one-point AG codes
over the Miura-Kamiya $C_{ab}$ curves
proposed by \citet*{lee11}
to general one-point AG codes,
without any assumption.
We also extend their unique decoding algorithm
to list decoding, modify it so that it can be used with
the Feng-Rao improved code construction,
prove equality between its error correcting capability
and half the minimum distance lower bound by \citet{andersen08} that
has not been done
in the original proposal except for one-point
Hermitian codes, remove the unnecessary computational
steps so that it can run faster,
and analyze its computational complexity in terms of multiplications
and divisions in the finite field.
As a unique decoding algorithm,
the proposed one is empirically and theoretically
as fast as the BMS algorithm for
one-point Hermitian codes.
As a list decoding algorithm,
extensive experiments suggest that it
can be
  much faster for many moderate size/usual inputs
  than the algorithm by \citet{beelen10}.
It should be noted that
  as a list decoding algorithm the proposed method seems to have
  exponential worst-case computational complexity while the previous
  proposals \citep{beelen10,guruswami99} have polynomial ones,
  and that the proposed method is expected to be slower than
  the previous proposals for very large/special inputs.
\end{abstract}

\section{Introduction}
We consider the list decoding of one-point algebraic geometry (AG) codes.
\citet{guruswami99} proposed the well-known
list decoding
algorithm for one-point AG codes, which consists of
the interpolation step and the factorization step.
The interpolation step has large computational complexity and
many researchers
have proposed faster interpolation steps,
see \citet[Figure 1]{beelen10}.

By modifying the unique decoding algorithm \citep{lee11}
for primal one-point AG codes,
we propose another list decoding algorithm
based on voting in Gr\"obner bases
whose error correcting capability is higher than
\citet{guruswami99} and whose computational
complexity is empirically
smaller than \citet{beelen10,guruswami99}
in many cases that are examined and reported in our
  computational experiments.
It should be noted that
  as a list decoding algorithm the proposed method seems to have
  exponential worst-case computational complexity while the previous
  proposals \citep{beelen10,guruswami99} have polynomial ones,
  and that the proposed method is expected to be slower than
  the previous proposals for very large/special inputs.
A decoding algorithm for primal one-point AG codes
was proposed in \citet{ldecodepaper}, which was a straightforward
adaptation of the original Feng-Rao majority voting for
the dual AG codes \citep{fengrao93} to the primal ones.
The Feng-Rao majority voting in \citet{ldecodepaper} for
one-point primal codes was generalized to multi-point primal
codes
in \citet[Section 2.5]{beelen08}.
The one-point primal codes can also be decoded as
multi-point dual codes with majority voting \citep{beelen07,duursma11,duursma10},
whose faster version was proposed in \citet{sakata11}
for the multi-point Hermitian codes.
\citet*{lee11}
proposed another unique decoding (not list decoding)
algorithm for primal codes based on the majority voting
inside Gr\"obner bases. The module used by them \citep{lee11}
is a curve theoretic generalization of one used for 
Reed-Solomon codes in \citet{kuijper11}
that is a special case of the module used
in \citet{lee08}.
An interesting feature in \citet{lee11} is that
it did not use differentials and residues on curves for its
majority voting, while they were used in \citet{beelen08,ldecodepaper}.
The above studies \citep{beelen08,lee11,ldecodepaper} dealt with
the primal codes.
We recently proved in
\citet{gmr13} that the error-correcting
capabilities of \citet{lee11,ldecodepaper} are the same.
The earlier papers \citep{duursma94,pellikaan93} suggest that
central observations in \citet{andersen08,gmr13,ldecodepaper} were known to the
Dutch group, which is actually the case \citep{duursma12pcomm}.
\citet{chen99},
\citet{elbrondjensen99}
and \citet{amoros06}
studied the error-correcting capability of the Feng-Rao \citep{fengrao93} or
the BMS algorithm \citep{sakata95b,sakata95a}
with majority voting beyond half the designed distance
that are applicable to the dual one-point codes.

There was room for improvements in the original
result \citep{lee11}, namely,
(a) they have not clarified
the relation between its error-correcting capability
and existing minimum distance lower bounds except
for the one-point Hermitian codes, 
(b) they have not analyzed the computational complexity,
(c) they assumed that the maximum
pole order used for code construction is less than the code length,
and (d) they have not shown how to use the method with the
Feng-Rao improved code construction \citep{feng95}.
We shall
(1) prove that the error-correcting capability of the original
proposal is always equal to half of the bound in
\citet{andersen08} for the minimum distance of one-point
primal codes (Proposition \ref{prop:AG}), (2)
generalize their algorithm to work with any one-point
AG codes, (3) modify their algorithm to a list decoding
algorithm, (4) remove the assumptions (c) and (d) above,
(5) remove unnecessary computational steps from the
original proposal,
(6) analyze the computational complexity in terms of the
number of multiplications and divisions in the finite field.
We remark that a generalization of \citet{lee11} to arbitrary primal
AG code is also reported in \citet{lee14} using a similar idea in
this paper reported earlier as a conference paper \citep{gmr12isit}.

The proposed algorithm is implemented on the Singular computer algebra
system \citep{singular313}, and we verified that
the proposed algorithm can correct more errors than 
\citet{beelen10,guruswami99} with manageable
computational complexity in many cases
  that are examined and reported in our computational
  experiments.

This paper is organized as follows:
Section \ref{sec2} introduces notations and relevant facts.
Section \ref{sec3} improves \citet{lee11} in various ways,
and the differences to the original \citep{lee11} are summarized
in Section \ref{sec:diff}.
Section \ref{sec4} shows that the proposed modification to
\citet{lee11} works as claimed.
Section \ref{sec:experiment} compares its computational
complexity with the conventional methods.
Section \ref{sec6} concludes the paper.
Part of this paper was presented
at 2012 IEEE International Symposium on Information Theory,
Cambridge, MA, USA, July 2012 \citep{gmr12isit}.

\section{Notation, Preliminaries and the Statement of Problem}\label{sec2}
Our study heavily relies on the standard form of algebraic curves
introduced independently by \citet{geilpellikaan00} and \citet{miura98},
which is an enhancement of earlier results \citep{miura92,saints95}.
Let $F/\mathbf{F}_q$ be an algebraic function field of one variable over
a finite field $\mathbf{F}_q$ with $q$ elements.
Let $g$ be the genus of $F$.
Fix $n+1$ distinct places $Q$, $P_1$, \ldots,
$P_n$ of degree one in $F$ and a nonnegative integer
$u$. We consider the following
one-point algebraic geometry (AG) code
\begin{equation}
C_u = \{ \ev(f) \mid f \in \mathcal{L}(uQ)\}
\label{eq:cu}
\end{equation}
where $\ev(f) = (f(P_1)$, \ldots, $f(P_n))$.

The problem considered in this paper is as follows:
We use a linear code $C_u$ (or its improved version $C_\Gamma$
defined in Eq.\ (\ref{eq:cgamma})).
A codeword in $C_u$ (or $C_\Gamma$) is transmitted,
and the receiver receives $\vec{r} \in \mathbf{F}_q^n$.
A list-decoding algorithm with a decoding radius $\tau$
finds all the codewords in $C_u$ (or $C_\Gamma$) whose
Hamming distances from $\vec{r}$ are within $\tau$.
Our aim is to propose another list-decoding algorithm
based on Gr\"obner bases and demonstrate its
computational complexity in a wide range of examples.

Suppose that the Weierstrass semigroup $H(Q)$
at $Q$ is generated by
$a_1$, \ldots, $a_t$, and choose
$t$ elements $x_1$, \ldots,
$x_t$ in $F$ whose pole divisors are $(x_i)_\infty = a_iQ$ for $i=1$, \ldots, $t$.
We do \emph{not} assume that
$a_1$ is the smallest among $a_1$, \ldots, $a_t$.
Without loss of generality we may assume the availability of
such $x_1$, \ldots,
$x_t$, because otherwise we cannot find a basis of $C_u$ for every $u$.
Then we have that $\mathcal{L}(\infty Q) = \cup_{i=1}^\infty\mathcal{L}(iQ)$
is equal to $\mathbf{F}_q[x_1$, \ldots, $x_t]$ \citep{saints95}.
We express $\mathcal{L}(\infty Q)$ as a residue class ring
$\mathbf{F}_q[X_1$, \ldots, $X_t]/I$
of the polynomial ring $\mathbf{F}_q[X_1$, \ldots, $X_t]$, where
$X_1$, \ldots, $X_t$ are transcendental over $\mathbf{F}_q$,
and $I$ is the kernel of the canonical homomorphism sending
$X_i$ to $x_i$. \citet{geilpellikaan00,miura98}
identified the following convenient representation of
$\mathcal{L}(\infty Q)$ by using Gr\"obner basis theory
\citep{buchberger65}
(See, for example, a textbook \citep{adams94}).
The following review is borrowed from \citet{miuraform}.
Hereafter, we assume that the reader is familiar with the
Gr\"obner basis theory in \citet{adams94}.

Let $\mathbf{N}_0$ be the set of nonnegative integers.
For $(m_1$, \ldots, $m_t)$, $(n_1$, \ldots, $n_t) \in
\mathbf{N}_0^t$,
we define the weighted reverse lexicographic monomial order $\succ$
such that $(m_1$, \ldots, $m_t)$ $\succ$ $(n_1$, \ldots, $n_t)$
if $a_1 m_1 + \cdots + a_t m_t > a_1 n_1 + \cdots + a_t n_t$,
or $a_1 m_1 + \cdots + a_t m_t = a_1 n_1 + \cdots + a_t n_t$,
and $m_1 = n_1$, $m_2 = n_2$, \ldots,
$m_{i-1} = n_{i-1}$, $m_i<n_i$, for some $1 \leq i \leq t$.
Note that a Gr\"obner basis of $I$ with respect to $\succ$
can be computed by \citet[Theorem 15]{saints95}, \citet{schicho98},
\citet[Theorem 4.1]{tang98} or
\citet[Proposition 2.17]{bn:vasconcelos},
starting from any affine defining equations of $F/\mathbf{F}_q$.

\begin{exmp}\label{ex1}
The Klein quartic over $\mathbf{F}_8$ is given by the equation
\[
u^3 v + v^3 + u = 0.
\]
There exists a unique $\mathbf{F}_8$-rational place $Q$,
namely $(0:1:0)$, that is a unique pole of $v$.
The numbers $3$, $5$ and $7$ is the minimal
generating set of the Weierstrass semigroup at $Q$.
Define three functions
$x_1 = v$, $x_2 = uv$,
$x_3 = u^2v$.
We have $(x_1)_\infty = 3Q$,
$(x_2)_\infty = 5Q$,
$(x_3)_\infty = 7Q$ as shown in \citet[Example 3.7]{hoholdt95}.
By \citet[Theorem 4.1]{tang98}
we can see that the standard form of the Klein quartic is
given by
\[
X_2^2+X_3X_1, X_3X_2+X_1^4+X_2, X_3^2+X_2X_1^3+X_3,
\]
which is the reduced Gr\"obner basis with respect to
the monomial order $\succ$.
We can see that $a_1=3$, $a_2=5$, and $a_3=7$.
\end{exmp}

\begin{exmp}\label{ex:gs1}
Consider the function field
$\mathbf{F}_9(u_1,v_2,v_3)$ with relations
\begin{equation}
v_2^3+v_2 = u_1^4, \quad v_3^3 + v_3 = (v_2/u_1)^4. \label{eq:gs1}
\end{equation}
This is the third function field in the asymptotically
good tower introduced by \citet{garcia95}.
Substituting $v_2$ with $u_1u_2$ and $v_3$ with $u_2 u_3$ in
Eq.\ (\ref{eq:gs1}) we have affine defining equations
\[
u_1^2u_2^3+u_2-u_1^3=0, \quad u_2^2u_3^3+u_3-u_2^3=0.
\]
 in $\mathbf{F}_9(u_1,u_2,u_3)
=\mathbf{F}_9(u_1,v_2,v_3)$.
The minimal generating set of the Weiestrass semigroup $H(Q)$ at $Q$
is $9$, $12$, $22$, $28$, $32$ and $35$ \citep[Example 4.11]{voss97}.
It has genus $22$ and
$77$ $\mathbf{F}_9$-rational points different from
$Q$ \citep{garcia95}.

Define six functions
$x_1 = u_1$, $x_2 = u_1 u_2$,
$x_3 = u_1^2u_2u_3$,
$x_4 = u_1^3u_2^2u_3^2$,
$x_5 = ((u_1u_2)^2+1)u_2u_3$ and
$x_6 = ((u_1u_2)^2+1)u_2^2u_3^2$.
We have $(x_1)_\infty = 9Q$,
$(x_2)_\infty = 12Q$,
$(x_3)_\infty = 22Q$,
$(x_4)_\infty = 35Q$,
$(x_5)_\infty = 28Q$ and
$(x_6)_\infty = 32Q$ \citep{umehara98}. From
this information and \citet[Theorem 4.1]{tang98} we can compute the $15$ polynomials
in the reduced Gr\"obner basis
of the ideal $I \subset \mathbf{F}_9[X_1$, \ldots, $X_6]$ defining
$\mathcal{L}(\infty Q)$ as
$\{X_2^3-X_1^4+X_2$,
$X_5X_2-X_3X_1^2$,
$X_6X_2-X_4X_1$,
$X_3^2-X_4X_1$,
$X_3X_2^2-X_5X_1^2+X_3$,
$X_5X_3-X_6X_1^2$,
$X_6X_3-X_1^6+X_5X_1^2+X_2X_1^2$,
$X_5^2-X_4X_2X_1-X_6$,
$X_4X_3-X_2X_1^5+X_3X_1^3+X_2^2X_1$,
$X_4X_2^2-X_6X_1^3+X_4$,
$X_6X_5-X_2^2X_1^4+X_3X_2X_1^2+X_5$,
$X_5X_4-X_1^7+X_5X_1^3+X_2X_1^3$,
$X_6^2-X_5X_1^4+X_4X_2X_1+X_3X_1^2+X_6$,
$X_6X_4-X_3X_1^5+X_6X_1^3+X_3X_2X_1$,
$X_4^2-X_3X_2X_1^4+X_4X_1^3+X_5X_1^2-X_3\}$.
Note that polynomials in the above Gr\"obner basis are
in the ascending order with respect to the monomial order $\prec$
while terms in each polynomial are
in the descending order with respect to $\prec$.
\end{exmp}

For $i=0$, \ldots, $a_1-1$,
we define $b_i = \min\{ m \in H(Q) \mid m \equiv i \pmod{a_1}\}$,
and $L_i$ to be the minimum element $(m_1$, \ldots,
$m_t) \in \mathbf{N}_0^t$
with respect to $\prec$ such that $a_1 m_1 + \cdots +
a_t m_t = b_i$.
Note that $b_i$'s are the well-known Ap\'ery set \citep[Lemmas 2.4 and 2.6]{MR2549780} of the numerical semigroup $H(Q)$.
Then we have $\ell_1 = 0$ if we write
$L_i$ as $(\ell_1$, \ldots, $\ell_t)$.
For each $L_i = (0$, $\ell_{i2}$, \ldots, $\ell_{it})$,
define $y_i = x_2^{\ell_{i2}} \cdots x_t^{\ell_{it}} \in \mathcal{L}(\infty Q)$.

The footprint of $I$, denoted by $\Delta(I)$,
is $\{ (m_1$, \ldots, $m_t) \in \mathbf{N}_0^t \mid X_1^{m_1} \cdots
X_t^{m_t}$ is not the leading monomial of
any nonzero polynomial in $I$ with respect to $\prec\}$,
and define $\Omega_0 = \{x_1^{m_1} \cdots x_t^{m_t} \mid 
(m_1$, \ldots, $m_t) \in \Delta(I)\}$.
Then $\Omega_0$ is a basis of $\mathcal{L}(\infty Q)$
as an  $\mathbf{F}_q$-linear space \citep{adams94},
two distinct elements in $\Omega_0$ have different pole orders at $Q$,
and
\begin{eqnarray}
\Omega_0 &=& \{ x_1^m x_2^{\ell_2} \cdots, x_t^{\ell_t} \mid m \in \mathbf{N}_0,
(0, \ell_2, \ldots, \ell_t) \in \{L_0, \ldots, L_{a_1-1}\}\}\nonumber\\
&=& \{ x_1^m y_i \mid m \in \mathbf{N}_0, i=0, \ldots, a_1-1\}.
\label{eq:footprintform}
\end{eqnarray}
Equation (\ref{eq:footprintform}) shows that
$\mathcal{L}(\infty Q)$ is a free $\mathbf{F}_q[x_1]$-module
with a basis $\{y_0$, \ldots, $y_{a_1-1}\}$.
Note that the above structured shape of $\Omega_0$ reflects the well-known
property of every weighted reverse lexicographic monomial order,
see the paragraph preceding to \citet[Proposition 15.12]{bn:eisenbud}.

\begin{rem}
For a description of the proposed algorithm,
only $a_1$, $b_1$, \ldots, $b_{a_1-1}$,
$y_1$, \ldots, $y_{a_1-1}$ are absolutely necessary,
and we can remove $\varphi_s$,
$x_2$, \ldots, $x_t$, $a_2$, \ldots, $a_t$, $X_1$, \ldots, $X_t$,
$Y_1$, \ldots, $Y_t$ from our presentation.

However, we retain those notations for the following two reasons:
Firstly, because $s$ will be used as the iteration variable
and our proposed algorithm will focus on the monomial
in $\Omega_0$ whose pole order at $Q$ is $s$,
the notation $\varphi_s$ clarifies the relation between $s$
and the monomial in $\Omega_0$ focused in the $s$-th iteration.

Secondly, the proposed algorithm needs to find the normal
form of a function in $\mathcal{L}(\infty Q)$ that is
an $\mathbf{F}_q$-linear combination of monomials in $\Omega_0$.
Functions in $\mathcal{L}(\infty Q)$ are represented by
multivariate polynomials over $\mathbf{F}_q$.
The specific definition of $X_1$, \ldots, $X_t$ and
the specific choice of the monomial order in
$\mathbf{F}_q[X_1$, \ldots, $X_t]$ enable us to
compute the normal form just by the standard Gr\"obner basis
division, and make implementation of our proposal easier
in practice.
Thirdly, $Y_1$, \ldots, $Y_t$ will be used to explain
how we count the computational cost.
They are necessary to allow readers to reproduce our
experimental results presented later in this paper.
\end{rem}

\begin{exmp}\label{ex2}
For the curve in Example \ref{ex1},
we have $y_0 = 1$, $y_1=x_3$, $y_2=x_2$.
\end{exmp}

Let $v_Q$ be the unique valuation in $F$ associated with the place $Q$. The semigroup $H(Q)$ is equal to $\{i a_1 - v_Q (y_j) \mid 0\le i,0\le j<a_1\}$
\citep[Lemma 2.6]{MR2549780}.
By \citet[Proposition 3.18]{miuraform}, for each nongap $s\in H(Q)$
there is a unique monomial $x_1^i y_j \in \Omega_0$ with $0\le j<a_1$ such
that $-v_Q (x_1^i  y_j )=s$, and let us denote this monomial
by $\varphi_s$. Let $\Gamma \subset H(Q)$, and we may consider
the one-point codes
\begin{equation}
C_\Gamma = \langle \{\ev(\varphi_s) \mid s \in \Gamma \}\rangle,\label{eq:cgamma}
\end{equation}
where $\langle \cdot \rangle$
denotes the $\mathbf{F}_q$-linear space spanned by $\cdot$.
Since considering linearly dependent rows in a generator matrix has no merit,
we assume
\begin{equation}
\Gamma \subseteq \widehat{H}(Q), \label{eq:gammaassume}
\end{equation}
where $\widehat{H}(Q) = \{ u \in H(Q) \mid
C_u \neq C_{u-1} \}$.
One motivation for considering these codes is that it was shown in
\citet{andersen08} how to increase the dimension of the one-point codes without decreasing the lower bound $d_{\mathrm{AG}}$
for the minimum distance.
The bound $d_{\mathrm{AG}}(C_\Gamma)$ is defined for $C_\Gamma$
as follows \citep{andersen08}:
For $s \in \Gamma$, let
\begin{equation}\lambda(s) = \sharp \{ j \in H(Q) \mid j+s \in
\widehat{H}(Q) \}. \label{eq:lambda}
\end{equation}
Then $d_{\mathrm{AG}}(C_\Gamma) = \min\{ \lambda(s) \mid s\in \Gamma \}$.
It is proved in \citet{geil11} that $d_{\mathrm{AG}}$ gives the
same estimate for the minimum distance as the Feng-Rao bound
\citep{fengrao93} for one-point
dual AG codes when both $d_{\mathrm{AG}}$ and the Feng-Rao
bound can be
applied, that is, when the dual of a one-point code is
isometric to a one-point code. Furthermore,
it is also proved in \citet{geil11} that
$d_{\mathrm{AG}}(C_\Gamma)$ can be obtained from the bounds in
\citet{beelen07,duursma11,duursma10}, hence $d_{\mathrm{AG}}$ can be
understood as a particular case of these bounds \citep{beelen07,duursma11,duursma10}.

\section{Procedure of New List Decoding based on Voting in Gr\"obner Bases}\label{sec3}
\subsection{Overall Structure}
Suppose that we have a received word $\vec{r} \in \mathbf{F}_q^n$.
We shall modify the unique decoding algorithm proposed by
\citet{lee11} so that we can find all the codewords
in $C_\Gamma$ in Eq.\ (\ref{eq:cgamma}) within the Hamming distance
$\tau$ from $\vec{r}$.
$\tau$ is a parameter independent of $\vec{r}$,
and $\tau$ is chosen before reception of $\vec{r}$.
The overall structure of the modified algorithm is as follows:
\begin{enumerate}
\item\label{l1} Precomputation before getting a received word $\vec{r}$,
\item\label{l2} Initialization after  getting a received word  $\vec{r}$,
\item\label{l3} Termination criteria of the iteration, and
\item\label{l4} Main part of the iteration.
\end{enumerate}
Steps \ref{l2} and \ref{l4} are based on \citet{lee11}.
Steps \ref{l1} and \ref{l3} are not given in \citet{lee11}.
Each step is described in the following subsections in
Section \ref{sec3}.
We shall analyze time complexity
except the precomputation part of the algorithm.

\subsection{Modified Definitions for the Proposed Modification}
We retain notations from Section \ref{sec2}.
In this subsection,
we modify notations and definitions in \citet{lee11}
to describe the proposed modification to their algorithm.
We also introduce several new notations.
Define a set
$\Omega_1 = \{ x_1^i y_jz^k \mid 0 \leq i$, 
$0 \leq j < a_1$, $k=0,1\}$.
Our $\Omega_1$ is $\Omega$ in \citet{lee11}.
Recall also that $\Omega_0 = \{ \varphi_s \mid s \in H(Q)\}$.

Since the $\mathbf{F}_q[x_1]$-module
$\RR z \oplus \RR$ has a free basis $\{ y_j z, y_j \mid
0 \leq j < a_1\}$,
we can regard $\Omega_1$ as the set of monomials in
the Gr\"obner basis theory for modules.
We introduce a monomial order on $\Omega_1$ as follows.
For given two monomials
$x_1^{i_1} y_j z^{i_{t+1}}$ and $x_1^{i'_1} y_{j'} z^{i'_{t+1}}$,
first rewrite $y_j$ and $y_{j'}$ by $x_2$, \ldots, $x_t$ defined
in Section \ref{sec2} and
get $x_1^{i_1} y_j z^{i_{t+1}} = x_1^{i_1} x_2^{i_2} \cdots x_t^{i_t} z^{i_{t+1}}$
and
$x_1^{i'_1} y_{j'} z^{i'_{t+1}} = x_1^{i'_1} x_2^{i'_2} \cdots x_t^{i'_t} z^{i'_{t+1}}$.
For a nongap $s \in H(Q)$,
we define the monomial order $x_1^{i_1} x_2^{i_2} \cdots x_t^{i_t} z^{i_{t+1}}<_s
x_1^{i'_1} x_2^{i'_2} \cdots x_t^{i'_t} z^{i'_{t+1}}$ parametrized by $s$
if $i_{t+1}s - v_Q(x_1^{i_1} x_2^{i_2} \cdots x_t^{i_t}) <
i'_{t+1}s - v_Q(x_1^{i'_1} x_2^{i'_2} \cdots x_t^{i'_t})$
or $i_{t+1}s - v_Q(x_1^{i_1} x_2^{i_2} \cdots x_t^{i_t}) =
i'_{t+1}s - v_Q(x_1^{i'_1} x_2^{i'_2} \cdots x_t^{i'_t})$
and $i_1 = i'_1$, $i_2 = i'_2$, \ldots, $i_{\ell-1}=i'_{\ell-1}$
and $i_\ell > i'_\ell$ for some $1 \leq \ell \leq t+1$.
Observe that the restriction of $<_s$ to $\Omega_0$ is
equal to $\prec$ defined in Section \ref{sec2}.
In what follows,
every Gr\"obner basis, leading term, and leading coefficient is
obtained by considering the Gr\"obner basis theory for modules,
not for ideals.

For $f \in \RR z \oplus \RR$,
$\gamma(f)$ denotes the number of nonzero terms in $f$
when $f$ is expressed as an $\mathbf{F}_q$-linear combination
of monomials in $\Omega_1$.
$\gamma_{\neq 1}(f)$ denotes the number of nonzero terms 
whose coefficients are not $1 \in \mathbf{F}_q$.

For the code $C_\Gamma$ in Eq.\ (\ref{eq:cgamma}),
define the divisor $D=P_1+ \cdots + P_n$.
Define $\mathcal{L}(-G+\infty Q) = \bigcup_{i=1}^\infty
\mathcal{L}(-G+iQ)$ for a positive
divisor $G$ of $F/\mathbf{F}_q$.
Then $\mathcal{L}(-D+\infty Q)$ is
an ideal of $\RR$ \citep{matsumoto99ldpaper}.
Let $\eta_i$ be any element in $\mathcal{L}(-D+\infty Q)$
such that $\textsc{lm}(\eta_i) = x_1^j y_i$ with $j$ being
the minimal given $i$.
Then by \citet[Proposition 1]{lee11},
$\{\eta_0$, \ldots, $\eta_{a_1-1}\}$ is a Gr\"obner basis
for $\mathcal{L}(-D+\infty Q)$ with respect to $<_s$
as an $\mathbf{F}_q[x_1]$-module.
For a nonnegative integer $s$, define
$\Gamma^{(\leq s)} = \{
s' \in \Gamma \mid s' \leq s \}$,
$\Gamma^{(> s)} = \{
s' \in \Gamma \mid s' > s \}$,
and $\mathrm{prec}(s)= \max\{s' \in H(Q) \mid s'<s \}$.
We define $\mathrm{prec}(0) = -1$.

\subsection{Precompuation before Getting a Received Word}\label{sec:precomput}
Before getting $\vec{r}$,
we need to compute the Pellikaan-Miura standard form of 
the algebraic curve, $y_0(=1)$, $y_1$, \ldots, $y_{a_1-1}$,
and $\varphi_s$ for $s \in H(Q)$ as defined in Section \ref{sec2}.
Also compute $\eta_0$, \ldots, $\eta_{a_1-1}$, which can be
done by \citet{matsumoto99ldpaper}.

For each $(i,j)$, express $y_i y_j$ as an $\mathbf{F}_q$-linear
combination of monomials in $\Omega_0$.
Such expressions will be used for computing products and quotients
in $\RR$ as explained in Section \ref{sec:product}.
{From} the above data, we can easily know
$\textsc{lc}(y_i y_j)$, which will be used in 
Eqs.\ (\ref{eq:quotientpre}) and (\ref{eq:mui}).

Find elements $\varphi_s \in \Omega_0$ with
$s \in \widehat{H}(Q)$.
There are $n$ such elements, which we denote by $\psi_1$,
\ldots $\psi_n$ such that $-v_Q(\psi_i) < -v_Q(\psi_{i+1})$.
Compute the $n \times n$ matrix
\begin{equation}
M=\left(
\begin{array}{ccc}
\psi_1(P_1)&\cdots&\psi_1(P_n)\\
\vdots&\vdots& \vdots\\
\psi_n(P_1)&\cdots&\psi_n(P_n)
\end{array}\right)^{-1}.\label{eq:invgenmatrix}
\end{equation}

\subsection{Multiplication and Division in an Affine Coordinate Ring}
In both of the original unique decoding algorithm \citep{lee11}
and our modified version,
we need to quickly compute the product $gh$
of two elements $g,h$ in the affine coordinate ring
$\RR$.
In our modified version, we also need to compute the quotient $g/h$
depending on the choice of iteration termination criterion
described in Section \ref{sec:termination}.
Since the authors could not find quick computational procedures
for those tasks in $\RR$, we shall present such ones here.

\subsubsection{Multiplication in an Affine Coordinate Ring}\label{sec:product}
The normal form of $g$, for $g \in \RR$, is the expression
of $g$ written as an $\mathbf{F}_q$-linear combination
of monomials $\varphi_s \in \Omega_0$.
$g,h$ are assumed to be in the normal form.
We propose the following procedure to compute the normal form of
$gh$. 
Let the normal form of $y_i y_j$ be
\[
\sum_{k=0}^{a_1-1} y_k f_{i,j,k}(x_1).
\]
with $f_{i,j,k}(x_1) \in \mathbf{F}_q[x_1]$,
which is computed in Section \ref{sec:precomput}.

We denote by $X_1$, $Y_1$, \ldots, $Y_{a_1-1}$
algebraically independent variables
over $\mathbf{F}_q$.
\begin{enumerate}
\item Assume that $g$ and $h$ are in their normal forms.
Change $y_i$ to $Y_i$ and $x_1$ to $X_1$ in $g,h$ for $i=1$, \ldots, $a_1-1$.
Recall that $y_0=1$.
Denote the results by $G,H$.
\item Compute $GH$.
This step needs
\begin{equation}
\gamma(g)\times \gamma(h)\label{eq:NFmulti1}
\end{equation}
multiplications in
$\mathbf{F}_q$.
\item Let $GH = \sum_{0\leq i,j <a_1} Y_i Y_j F_{G,H,i,j}(X_1)$.
Then we have
\begin{equation}
gh =  \sum_{0\leq i,j<a_1} F_{G,H,i,j}(x_1) \sum_{k=0}^{a_1-1} y_k f_{i,j,k}(x_1).
\label{eq:gh}
\end{equation}
Computation of $F_{G,H,i,j}(X_1)\sum_{k=0}^{a_1-1} y_k f_{i,j,k}(x_1)$ needs at most
$\gamma_{\neq 1}(\sum_{k=0}^{a_1-1} y_k f_{i,j,k}(x_1)) \allowbreak
\gamma( F_{G,H,i,j}(x_1))$ multiplications in $\mathbf{F}_q$.
Therefore, the total number of multiplications in $\mathbf{F}_q$ 
in this step is at most
\begin{equation}\sum_{0 \leq i,j < a_1} \gamma( F_{G,H,i,j}(x_1))
\gamma_{\neq 1}(\sum_{k=0}^{a_1-1} y_k f_{i,j,k}(x_1)).\label{eq:NFmulti2}
\end{equation}
\end{enumerate}
Therefore, the total number of multiplications in
$\mathbf{F}_q$ is at most
\begin{equation}
\gamma(g)\times \gamma(h) + \sum_{0 \leq i,j < a_1} \gamma( F_{G,H,i,j}(x_1))
\gamma_{\neq 1}(\sum_{k=0}^{a_1-1} y_k f_{i,j,k}(x_1))
. \label{eq:NFmulti3}
\end{equation}
Define Eq.\ (\ref{eq:NFmulti3}) as $\mathrm{multi}(g,h)$.

We emphasize  that when the characteristic of $\mathbf{F}_q$ is 2 and
all the coefficients of defining equations belong to $\mathbf{F}_2$,
which is almost always the case
for those cases of interest for applications in coding theory,
then
$\gamma_{\neq 1}(\sum_{k=0}^{a_1-1} y_k f_{i,j,k}(x_1))$ in Eq.\ (\ref{eq:NFmulti3}) is zero. This means that $\mathcal{L}(\infty Q)$ has little additional
overhead over $\mathbf{F}_q[X]$ for computing products of their elements
in terms of the number of $\mathbf{F}_q$-multiplications and divisions.

\begin{rem}
Define $(i,j)$ to be equivalent to $(i',j')$ if
$y_i y_j = y_{i'}y_{j'} \in \mathcal{L}(\infty Q)$.
Denote by $[i,j]$ the equivalence class represented by $(i,j)$.
For $(i,j), (i',j') \in [i,j]$
we have $f_{i,j,k}(x_1) = f_{i',j',k}(x_1)$, which is denoted by
$f_{[i,j],k}(x_1)$.
The right hand side of Eq.\ (\ref{eq:gh}) can be written as
\begin{equation}
\sum_{[i,j]}\left(\sum_{(i',j')\in[i,j]} F_{G,H,i',j'}(x_1)\right) \sum_{k=0}^{a_1-1} y_k f_{[i,j],k}(x_1). \label{eq:gh2}
\end{equation}
By using Eq.\ (\ref{eq:gh2}) instead of Eq.\ (\ref{eq:gh}),
we have another upper bound on the number of multiplications as
\begin{equation}
\gamma(g)\times \gamma(h) + \sum_{[i,j]} \gamma\left( \sum_{(i',j')\in[i,j]} F_{G,H,i',j'}(x_1)\right)
\gamma_{\neq 1}(\sum_{k=0}^{a_1-1} y_k f_{[i,j],k}(x_1))
. \label{eq:NFmulti4}
\end{equation}
Since
\[
\gamma\left( \sum_{(i',j')\in[i,j]} F_{G,H,i',j'}(x_1)\right)
\leq 
\sum_{(i',j')\in[i,j]} \gamma( F_{G,H,i',j'}(x_1)),
\]
we have Eq.\ (\ref{eq:NFmulti4}) $\leq$ Eq.\ (\ref{eq:NFmulti3}).
However, Eq.\ (\ref{eq:NFmulti4}) is almost always the same
as Eq.\ (\ref{eq:NFmulti3}) over the curve in Example \ref{ex:gs1},
and Eq.\ (\ref{eq:NFmulti4}) will not be used in our computer experiments
in Section \ref{sec:experiment}.
\end{rem}

\subsubsection{Computation of the Quotient}
Assume $h\neq 0$.
The following procedure computes the quotient $g/h \in \mathcal{L}(\infty Q)$
or declares that $g$ does not belong to the principal ideal of $\RR$
generated by $h$.
\begin{enumerate}
\item Initialize $\sigma=0$. Also initialize $\zeta=0$.
\item\label{lab} Check if $-v_Q(g) \in -v_Q(h) + H(Q)$. If not, declare
that $g$ does not belong to the principal ideal of $\RR$
generated by $h$, and finish the procedure.
\item Let $\varphi_s \in \Omega_0$ such that
$-v_Q(g) = -v_Q(\varphi_sh)$.
Observe that
$\textsc{lc}(\varphi_s\textsc{lm}(h)) = 
\textsc{lc}(y_{s \bmod a_1} y_{-v_Q(h) \bmod a_1})$ and that
$\textsc{lc}(y_{s \bmod a_1} y_{-v_Q(h) \bmod a_1})$ is precomputed
as Section \ref{sec:precomput}.
Let
\begin{equation}
\mathbf{F}_q \ni t =  \textsc{lc}(g)/(\textsc{lc}(h) \times \underbrace{\textsc{lc}(\varphi_s\textsc{lm}(h))}_{\textrm{Precomputed in Section \ref{sec:precomput}}}).
\label{eq:quotientpre}
\end{equation}
Computation of $t\varphi_s$ needs one multiplication and one division in $\mathbf{F}_q$.
Observe that $-v_Q(g - t\varphi_sh) < -v_Q(g)$.
\item Compute the normal form of $t \varphi_s h$, which requires at most
$\mathrm{multi}(t\varphi_s, h)$
multiplications in $\mathbf{F}_q$.
Increment $\zeta$ by $2+\mathrm{multi}(t\varphi_s, h)$.
\item Update $\sigma \leftarrow \sigma + t \varphi_s$ and $g \leftarrow g - t\varphi_sh$. If the updated $g$ is zero,
then output the updated $\sigma$ as the quotient and finish the procedure.
Otherwise go to Step \ref{lab}.
This step has no multiplication nor division.
\end{enumerate}
Define $\mathrm{quot}(g,h)$ as $\zeta$ after finishing the above
procedure. $\mathrm{quot}(g,h)$ is an upper bound on the number
of multiplications and divisions in $\mathbf{F}_q$ in the
above procedure. The program variable $\zeta$ is just to define
$\mathrm{quot}(g,h)$, and the decoding algorithm does not
need to update $\zeta$.
Observe also that the above procedure is a straightforward
generalization of the standard long division of two univariate polynomials \citep{mckeague11}.

\subsection{Initialization after  Getting a Received Word $\vec{r}$}\label{sec:initialization}
Let $(i_1$, \ldots, $i_n)^T = M\vec{r}$, where
$M$ is defined in Eq.\ (\ref{eq:invgenmatrix}).
Define $h_{\vec{r}} = \sum_{j=1}^n i_j \psi_j$.
Then we have $\ev(h_{\vec{r}}) = \vec{r}$.
The computation of $h_{\vec{r}}$ from $\vec{r}$
needs at most $n^2$ multiplications in $\mathbf{F}_q$.

Let $N=-v_Q(h_{\vec{r}})$.
For $i=0$, \ldots, $a_1-1$,
compute $g_i^{(N)} = \eta_i \in \RR$ and $f_i^{(N)} =
y_i(z-h_{\vec{r}}) \in \RR z \oplus \RR$.
The computation of $f_i^{(N)}$ needs
at most $\mathrm{multi}(y_i, h_{\vec{r}})$ multiplications in
$\mathbf{F}_q$. Therefore, the total number of multiplications
in the initialization is at most
\begin{equation}
n^2 + \sum_{i=0}^{a_1-1} \mathrm{multi}(y_i, h_{\vec{r}}). \label{eq:compl:ini}
\end{equation}
Let $s=N$ and execute the following steps.

\subsection{Three Termination Criteria of the Iteration}\label{sec:termination}
After finishing the initialization step in Section \ref{sec:initialization},
we iteratively compute $f_i^{(s)}$ and $g_i^{(s)}$ with $N \geq s \in H(Q) \cup
\{-1\}$ and $w_s$ with $N \geq s \in H(Q)$
 from larger $s$ to smaller $s$.
The single iteration consists of two parts:
The first part is to check if an iteration termination criterion
is satisfied.
The second part is computation of $f_i^{(s)}$ and $g_i^{(s)}$
for $N \geq s \in H(Q) \cup \{-1\}$.
We describe the first part in Section \ref{sec:termination}.

Let $f_{\mathrm{min}} = \alpha_0 + z \alpha_1$ having the smallest $-v_Q(\alpha_1)$ among $f^{(s)}_0$, \ldots, $f^{(s)}_{a_1-1}$.
In the following subsections, we shall propose three different
procedures to judge whether or not iterations in the proposed algorithm
can be terminated.
In an actual implementation of the proposed algorithm,
one criterion is chosen and the chosen one is consistently used
throughout the iterations.
The first one and the second one are different
generalizations
of \citet[Theorem 12]{kuijper11} from the case $g=0$ to
$g > 0$.
\citet[Theorem 12]{kuijper11} proved
that if the number $\delta$ of errors satisfies
$2 \delta < d_{\mathrm{RS}}(C_s)$,
where $d_{\mathrm{RS}}(C_s)$ is the minimum distance
$n-s$ of the $[n,s+1]$ Reed-Solomon code $C_s$,
then the transmitted information word is obtained by Ali-Kuijper's algorithm
as $-\alpha_0/\alpha_1$.
To one-point primal AG codes,
$d_{\mathrm{RS}}(C_s)$ can be generalized 
as
either $d_{\mathrm{AG}}(C_s)$ or  $n-s-g$.
The former generalization $d_{\mathrm{AG}}(C_s)$ corresponds to 
the first criterion in Section \ref{sec:first}
and the latter $n-s-g$ corresponds to the second in Section \ref{sec:second}.

The third one is almost the same
as the original procedure in \citet{lee11}.
The first one was proposed in \citet{gmr12isit} while
the second and the third ones are new in this paper.
We shall compare the three criteria in Section \ref{sec:comparethree}.
Throughout this paper,
$\wt(\vec{x})$ denotes the Hamming weight of a vector $\vec{x}\in \mathbf{F}_q^n$.

The proposed decoding algorithm can perform both
unique decoding and list decoding depending on
whether or not $d_{\mathrm{AG}}(C_{\Gamma^{(\leq s)}}) > 2 \tau$.
To provide a unified presentation of both variants of
our proposal, we will include the condition $d_{\mathrm{AG}}(C_{\Gamma^{(\leq s)}}) > 2 \tau$ in the following.
Observe that the condition $d_{\mathrm{AG}}(C_{\Gamma^{(\leq s)}}) > 2 \tau$
can be checked when one implements our proposal by 
an electronic circuit or a computer software,
before executing the proposed algorithm.

\subsubsection{First Criterion for Judging Termination}\label{sec:first}
If
\begin{itemize}
\item $s \in \Gamma$,
\item $d_{\mathrm{AG}}(C_{\Gamma^{(\leq s)}}) > 2 \tau$, and
\item $-v_Q(\alpha_1) \leq \tau + g$
\end{itemize}
then do the following:
\begin{enumerate}
\item\label{first1} Compute $\alpha_0/\alpha_1 \in F$. This needs at most
\begin{equation}\mathrm{quot}(\alpha_0, \alpha_1)\label{eq:compl3}
\end{equation}
multiplications and divisions
in $\mathbf{F}_q$.
\item If $\alpha_0/\alpha_1 \in \RR$ and $\alpha_0/\alpha_1$
can be written as a linear combination of monomials
in $\{\varphi_{s'}\in s'\in \Gamma^{(\leq s)}\}$, then do the following:
\begin{enumerate}
\item\label{first11} If $d_{\mathrm{AG}}(C_{\Gamma}) > 2 \tau$ or
$-v_Q(\alpha_1) \leq \tau$ then
include the coefficients of $-\alpha_0/\alpha_1+\sum_{s' \in \Gamma^{(> s)}} w_{s'} \varphi_{s'}$ into the list of transmitted
information vectors, and avoid proceeding with
the rest of the decoding procedure.
When $\Gamma^{(> s)} = \emptyset$, the summation over $\Gamma^{(> s)}$ is regarded as zero.
\item\label{first12} Otherwise  compute
$\ev(-\alpha_0/\alpha_1 +\sum_{s' \in \Gamma^{(> s)}} w_{s'} \varphi_{s'})$.
This needs at most
\begin{equation}
n \gamma(-\alpha_0/\alpha_1+\sum_{s' \in \Gamma^{(> s)}} w_{s'} \varphi_{s'}) \label{eq:compl4}
\end{equation}
multiplications and divisions
in $\mathbf{F}_q$.
\item\label{first13} If
\[ \wt \left(\ev(-\alpha_0/\alpha_1+\sum_{s' \in \Gamma^{(> s)}} w_{s'} \varphi_{s'})- \vec{r}\right) \leq  \tau,
\]
then include the coefficients of $-\alpha_0/\alpha_1$ $+$ $\sum_{s' \in \Gamma^{(> s)}} w_{s'} \varphi_{s'}$ into the list of transmitted
information vectors, and avoid proceeding with
$s$. Otherwise, continue the iterations
unless $s < n - g - 2\tau$.
\end{enumerate}
\end{enumerate}

\subsubsection{Second Criterion for Judging Termination}\label{sec:second}
If $s = \max\{ s' \in \Gamma 
\mid s' < n-2 \tau - g\}$,
then do the following:
\begin{enumerate}
\item\label{second1} If $-v_Q(\alpha_1) > \tau + g$ then stop proceeding with 
iteration. 
\item Otherwise compute $\alpha_0/\alpha_1 \in F$. This needs at most
\begin{equation}
\mathrm{quot}(\alpha_0, \alpha_1) \label{eq:compl5}
\end{equation}
multiplications and divisions
in $\mathbf{F}_q$.
\item If $2 \tau < d_{\mathrm{AG}} (C_\Gamma )$,
$\alpha_0/\alpha_1 \in \RR$, and $\alpha_0/\alpha_1$
can be written as a linear combination of monomials
in $\{\varphi_{s'}\in s'\in \Gamma^{(\leq s)}\}$
then
declare the coefficients of $-\alpha_0/\alpha_1+\sum_{s' \in \Gamma^{(> s)}} w_{s'} \varphi_{s'}$ as the only transmitted information and finish.
When $\Gamma^{(> s)} = \emptyset$, the summation over $\Gamma^{(> s)}$ is regarded as zero.
Otherwise declare ``decoding failure'' and finish.
\item If $2 \tau \geq d_{\mathrm{AG}} (C_\Gamma )$,
$\alpha_0/\alpha_1 \in \RR$ and $\alpha_0/\alpha_1$
can be written as a linear combination of monomials
in $\{\varphi_{s'}\in s'\in \Gamma^{(\leq s)}\}$, then  do the following:
\begin{enumerate}
\item\label{second11} If $-v_Q(\alpha_1) \leq \tau$ then 
include the coefficients of $-\alpha_0/\alpha_1+\sum_{s' \in \Gamma^{(> s)}} w_{s'} \varphi_{s'}$ into the list of transmitted
information vectors, and avoid proceeding with
$s$.
\item\label{second12} Otherwise compute
$\ev(-\alpha_0/\alpha_1 +\sum_{s' \in \Gamma^{(> s)}} w_{s'} \varphi_{s'})$.
This needs at most
\begin{equation}
n \gamma(-\alpha_0/\alpha_1+\sum_{s' \in \Gamma^{(> s)}} w_{s'} \varphi_{s'})\label{eq:compl6}
\end{equation}
multiplications and divisions
in $\mathbf{F}_q$.
\item If
\[ \wt \left(\ev(-\alpha_0/\alpha_1+\sum_{s' \in \Gamma^{(> s)}} w_{s'} \varphi_{s'})- \vec{r}\right) \leq  \tau,
\]
then include the coefficients of $-\alpha_0/\alpha_1$ $+$ $\sum_{s' \in \Gamma^{(> s)}} w_{s'} \varphi_{s'}$ into the list of transmitted
information vectors.
\end{enumerate}
\item Finish the iteration no matter what happened in the above steps.
\end{enumerate}

\subsubsection{Third Criterion for Judging Termination}\label{sec:3rd}
The third criterion terminates the algorithm when
it finds $f_i^{(s)}$ at $s=-1$ for $i=0$, \ldots, $a_1-1$
in the main iteration presented in Section \ref{sec:mainiteration}.
If $2 \tau < d_{\mathrm{AG}} (C_\Gamma )$ then
declare the vector $(w_s : s \in \Gamma)$ as the only transmitted information and finish.

If $2 \tau \geq d_{\mathrm{AG}} (C_\Gamma )$ then do the following:
\begin{enumerate}
\item\label{third1} If $\alpha_0=0$ and $-v_Q(\alpha_1) \leq \tau$ then
include the vector $(w_s : s \in \Gamma)$  into the list of transmitted
information vectors. Finish the iteration.
\item\label{third2} If $-v_Q(\alpha_1) > \tau+g$ then
finish the iteration.
\item\label{third3} Otherwise compute
$\ev(\sum_{s \in \Gamma} w_{s} \varphi_{s})$.
This needs at most
\begin{equation}
n \gamma(\sum_{s \in \Gamma} w_{s} \varphi_{s})\label{eq:compl7}
\end{equation}
 multiplications and divisions
in $\mathbf{F}_q$.
\item\label{third4} If
\[ \wt \left(\ev(\sum_{s \in \Gamma} w_{s} \varphi_{s})- \vec{r}\right) \leq  \tau,
\]
then include the vector $(w_s : s \in \Gamma)$  into the list of transmitted
information vectors. Finish the iteration.
\end{enumerate}

\subsection{Iteration of Pairing, Voting, and Rebasing}\label{sec:mainiteration}
The iteration of the original algorithm \citep{lee11} consists of
three steps, called pairing, voting, and rebasing.
We will make a small change to the original.
Our modified version is described below.

\subsubsection{Pairing}
Let
\[
g_i^{(s)} = \sum_{0\le j<a_1}c_{i,j}  y_j  z+ \sum_{0\le j<a_1} d_{i,j}    y_j, \mathrm{~with~} c_{i,j}, d_{i,j} \in\mathbf{F}_q[x_1],
\]
\[
f_i^{(s)}   =   \sum_{0\le j<a_1}a_{i,j}   y_j  z+ \sum_{0\le j<a_1}b_{i,j}   y_j, \mathrm{~with~} a_{i,j}, b_{i,j} \in\mathbf{F}_q[x_1],
\]
and let $\nu_i^{(s)}=\textsc{lc} (d_{i,i})$.
We assume that $\textsc{lt}(f_i^{(s)}) = a_{i,i}   y_i  z$ and
$\textsc{lt}(g_i^{(s)}) = d_{i,i}   y_i$.
For $0 \le i <a_1$, as in \citet{lee11}
there are unique integers $0\le i'<  a_1$ and $k_i$ satisfying
\[
-v_Q(a_{i,i} y_i) + s = a_1 k_i -v_Q(y_{i'}).
\]
Note that by the definition above
\begin{equation}\label{eq:iprime}i' = i + s \bmod  a_1,\end{equation}
and the integer $-v_Q (a_{i,i} y_i )+s$ is a nongap if and only if $k_i\ge 0$. Now let $c_i=\deg_{x_1}(d_{i',i'})-k_i$. Note that the map $i\mapsto i'$ is a permutation of $\{0,1,\dots,a-1\}$ and that  the integer $c_i$ is defined such that $a_1 c_i= -v_Q (d_{i',i'}  y_{i'} )+v_Q (a_{i,i}  y_i )-s$.

\subsubsection{Voting}
For each $i \in  \{0, \ldots a_1 -1\}$, we set
\begin{equation}
\mu_i= \textsc{lc}( a_{i,i} y_i \varphi_s), ~ w_{s,i}=-\frac{b_{i,i'}[x_1^{k_i}]}{\mu_i}, ~  \bar{c}_i=\max\{c_i,0\},
\label{eq:mui}
\end{equation}
where $b_{i,i'}[x_1^{k_i}]$ denotes the coefficient of
$x_1^{k_i}$ of
the univariate polynomial $b_{i,i'} \in \mathbf{F}_q[x_1]$.
We remark that the leading coefficient $\mu_i$ must be
considered after expressing $a_{i,i} y_i \varphi_s$ by monomials in $\Omega_0$.

Observe that $\textsc{lc}(y_i \varphi_s) = \textsc{lc}(y_i y_{s \bmod a_1})$
and that $\textsc{lc}(y_i y_{s \bmod a_1})$ is already precomputed
as Section \ref{sec:precomput}.
By using that precomputed table,
computation of $\mu_i$ needs one multiplication.
The total number of multiplications and divisions in Eq.\ (\ref{eq:mui}) is
\begin{equation}
2 a_1 \label{eq:compl1}
\end{equation}
excluding negation from the number of multiplication.

Let
\begin{equation}\nu(s)=\frac{1}{a_1}\sum_{0\le i<a_1}
\max\{-v_Q(\eta_{i'})+v_Q(  y_i )-s,0\}.\label{eq:nu}\end{equation}
We consider two different candidates depending on whether $s \in \Gamma$ or not:
\begin{itemize}
\item If $s \in H(Q) \setminus \Gamma$, set 
\begin{equation}w=0. \label{eq:w0}
\end{equation}
\item If $s \in \Gamma$, let $w$ be one of the element(s)
in $\mathbf{F}_q$ with
\begin{equation}\label{eq:acceptedvote}
	\sum_{w=w_{s,i}}\bar{c}_i \geq \sum_{w\neq w_{s,i}}\bar{c}_i - 2\tau+\nu(s).
\end{equation}
\end{itemize}
Let $w_s=w$. If several $w$'s satisfy the condition above,
repeat the rest of the algorithm for each of them.

\subsubsection{Rebasing}\label{sec:rebasing}
In all of the following cases,
we need to compute the normal form of the product
$w\varphi_s \times \sum_{j=0}^{a_1-1} a_{i,j}   y_j$,
and the product $w\varphi_s \times \sum_{j=0}^{a_1-1} c_{i,j}   y_j$.
For each $i$,
the number of multiplications is
\begin{equation}
\leq \mathrm{multi}(w\varphi_s, \sum_{j=0}^{a_1-1} a_{i,j}   y_j)
+ \mathrm{multi}(w\varphi_s, \sum_{j=0}^{a_1-1} c_{i,j}   y_j), \label{eq:multirebasing}
\end{equation}
where $\mathrm{multi}(\cdot, \cdot)$ is
defined in Section \ref{sec:product}.

\begin{itemize}
\item If $w_{s,i}=w$, then let
\begin{eqnarray*}\label{fkfmvd}
	&&g_{i'}^{( \mathrm{prec}(s))}=g_{i'}^{(s)}(z+w\varphi_s),\\
	&&f_i^{( \mathrm{prec}(s))}=f_i^{(s)}(z+w\varphi_s)
\end{eqnarray*}where the parentheses denote substitution of the variable $z$ and let $\nu_{i'}^{(  \mathrm{prec}(s))}=\nu_{i'}^{(s)}$. 
The number of multiplications in this case is bounded by Eq.\ (\ref{eq:multirebasing}).

\item If $w_{s,i}\neq w$ and $c_i>0$, then let
\begin{eqnarray*}
&&	  g_{i'}^{( \mathrm{prec}(s))}=f_i^{(s)}(z+w\varphi_s),  \\
&&	 f_i^{( \mathrm{prec}(s))}=x_1^{c_i}f_i^{(s)}(z+w\varphi_s)-\frac{\mu_i(w-w_{s,i}) }{\nu_{i'}^{(s)}}g_{i'}^{(s)}(z+w\varphi_s)
\end{eqnarray*}
and let $\nu_{i'}^{( \mathrm{prec}(s))}=\mu_i(w-w_{s,i})$.

Computation of $\frac{\mu_i(w-w_{s,i}) }{\nu_{i'}^{(s)}}$ needs 
one multiplication and one division.
The product of $\frac{\mu_i(w-w_{s,i}) }{\nu_{i'}^{(s)}}$ and
$g_{i'}^{(s)}(z+w\varphi_s)$ needs $\gamma(g_{i'}^{(s)}(z+w\varphi_s))$
multiplications, where $\gamma$ is defined in Section \ref{sec:product}.
Thus, the number of multiplications and divisions is
\begin{equation}
\leq 2 + \gamma(g_{i'}^{(s)}(z+w\varphi_s)) + \mbox{Eq.\ (\ref{eq:multirebasing})}. \label{eq:compl2}
\end{equation}

\item If $w_{s,i}\neq w$ and $c_i\le 0$, then let
\begin{eqnarray*}
&&	 g_{i'}^{({ \mathrm{prec}(s)})}=g_{i'}^{(s)}(z+w\varphi_s),\\
&& f_i^{({ \mathrm{prec}(s)})}=f_i^{(s)}(z+w\varphi_s)-\frac{\mu_i(w-w_{s,i})}{\nu_{i'}^{(s)}}x_1^{-c_i}g_{i'}^{(s)}(z+w\varphi_s)
\end{eqnarray*}
and let $\nu_{i'}^{( \mathrm{prec}(s))}=\nu_{i'}^{(s)}$.

Computation of $\frac{\mu_i(w-w_{s,i}) }{\nu_{i'}^{(s)}}$ needs 
one multiplication and one division.
The product of $\frac{\mu_i(w-w_{s,i}) }{\nu_{i'}^{(s)}}$ and
$g_{i'}^{(s)}(z+w\varphi_s)$ needs $\gamma(g_{i'}^{(s)}(z+w\varphi_s))$
multiplications, where $\gamma$ is defined in Section \ref{sec:product}.
Thus, the number of multiplications and divisions is
$\leq $ Eq.\ (\ref{eq:compl2}).
\end{itemize}
After computing $f_i^{(\mathrm{prec}(s))}$ and 
$g_i^{(\mathrm{prec}(s))}$ as above,
update the program variable $s$ to $\mathrm{prec}(s)$ and
return to the beginning of Section \ref{sec:termination}, that is, return to the step of the chosen termination criterion.

\subsection{Difference to the Original Method}\label{sec:diff}
In this subsection, we review advantages of our modified algorithm
over the original \citep{lee11}.
\begin{itemize}
\item Our version can handle any one-point primal AG codes,
while the original can handle codes only coming from the
$C_{ab}$ curves \citep{miura92}.
This generalization is enabled only by replacing $y^j$ in
\citet{lee11} by $y_j$ defined in Section \ref{sec2}.
\item Our version can find all the codewords within
Hamming distance $\tau$ from the received word $\vec{r}$,
while the original is a unique decoding algorithm.
\item Our version does not compute
$f^{(s)}_i$, $g^{(s)}_i$ for a Weiestrass gap $s \notin H(Q)$,
while the original computes them for $N\geq s\notin H(Q)$.
\item The original algorithm assumed $u<n$, where
$u$ is as defined in Eq.\ (\ref{eq:cu}). This assumption
is replaced by another less restrictive assumption (\ref{eq:gammaassume})
in our version.
\item Our version supports the Feng-Rao improved code
  construction \citep{feng95}, while the original does not.
This extension is made possible by the change at Eq.\ (\ref{eq:w0}).
\item The first and the second termination criteria come from
  \citet[Theorem 12]{kuijper11}
and do not exist in the original \citep{lee11}.
\item The third termination criterion is essentially the same as
  the original \citep{lee11}, but examination of
  the Hamming distance between the decoded codeword and $\vec{r}$
  is added when $2 \tau \geq d_{\mathrm{AG}}(C_\Gamma)$.
\item The original \citep{lee11} is suitable for parallel
  implementation on electric circuit similar to the K\"otter
architecture
\citep{koetter98}.
Our modified version retains this advantage.
\end{itemize}

\subsection{Small Example of Algorithm Execution}
We show a short example execution of the proposed list decoding.
Consider the well-known Hermitian function field
$\mathbf{F}_4(u,v)/\mathbf{F}_4$ with the relation
$v^2+v-u^3=0$, which implies $a_1 = 2$.
We choose the common pole of $u$ and $v$ as $Q$, and
$P_1=(0,0)$, $P_2=(0,1)$,
$P_3=(1,\beta)$,
$P_4=(1,\beta^2)$,
$P_5=(\beta,\beta)$,
$P_6=(\beta,\beta^2)$,
$P_7=(\beta^2,\beta)$,
$P_8=(\beta^2,\beta^2)$,
where $\beta$ is a primitive element in $\mathbf{F}_4$.
Then we have $\widehat{H}(Q) = \{0$, $2$, $3$, $4$, $5$, $6$, $7$, $9\}$.
The above choice gives
$\eta_0= u^4+u$ and $\eta_1 = u^4v+uv$.

We use $C_u$ with $u=4$ for decoding.
Then $\Gamma = \{0$, $2$, $3$, $4\}$
and $d_{\mathrm{AG}}(C_\Gamma) = 4$.
We shall try to correct $\tau=2$ errors.
Suppose that $\vec{0}$ was transmitted and
$\vec{r} = (0$, $0$, $1$, $1$, $0$, $0$, $0$, $0)$ was received.
This received word $\vec{r}$ has four codewords within
Hamming distance $2$, and all of them are found by the
proposed decoding algorithm. We shall show how one of the four
codeword is found.

According to Section \ref{sec:initialization},
we have $h_{\vec{r}}=u^3+u^2+u$, which implies $N=-v_Q(h_{\vec{r}}) = 6$.
We start the iteration from $s=6 (=N)$,
and again according to Section \ref{sec:initialization}
we have
\begin{eqnarray*}
f_0^{(6)}&=&z+u^3+u^2+u,\\
f_1^{(6)}&=&vz + u^3v+u^2v+uv,\\
g_0^{(6)}&=&u^4+u,\\
g_1^{(6)}&=&u^4v+uv.
\end{eqnarray*}

At the iteration for $s=6$ none of the temination criteria is satisfied.
Since $6 \notin \Gamma$, we do not need voting and set $w_6 = 0$,
and we compute new polynomials as
\begin{eqnarray*}
f_0^{(5)}&=&uz + u^3+u^2+u,\\
f_1^{(5)}&=&uvz+ u^3v +u^2v+uv,\\
g_0^{(5)}&=&z+u^3+u^2+u,\\
g_1^{(5)}&=&vz + u^3v+u^2v+uv.
\end{eqnarray*}

At the iteration for $s=5$ none of the temination criteria is satisfied.
Since $5 \notin \Gamma$, we do not need voting and set $w_5 = 0$,
and we compute new polynomials as
\begin{eqnarray*}
f_0^{(4)}&=&f_0^{(5)},\\
f_1^{(4)}&=&f_1^{(5)},\\
g_0^{(4)}&=&g_0^{(5)},\\
g_1^{(4)}&=&g_1^{(5)}.
\end{eqnarray*}

At the iteration for $s=4$ none of the temination criteria is satisfied.
We need voting at $s=4$, and all the candidates $0$, $1$, $\beta$,
$\beta^2$ in $\mathbf{F}_4$ get zero vote.
Therefore, the execution of the proposed algorithm splits to
four branches at $s=4$.
Firstly we consider the choice $w_4=1$, which will result in
an incorrect decoding.

The choice $w_4=1$ gives
\begin{eqnarray*}
f_0^{(3)}&=&uz+u^2+u,\\
f_1^{(3)}&=&uvz+u^2v+uv,\\
g_0^{(3)}&=&z+u^3+u,\\
g_1^{(3)}&=&vz+u^3v+uv.
\end{eqnarray*}

At the iteration for $s=3$ only the first
temination criterion is satisfied.
We have $\alpha_0 = u^2+u$ and $\alpha_1 = u$,
and the decoded codeword by the first criterion is
$(1$, $1$, $1$, $1$, $0$, $0$, $0$, $0)$ which is at the Hamming distance
$2$ from $\vec{r}$ but not equal to the transmitted codeword.

The second or the third termination criterion is not satisfied.
In the voting for $s=3$, $w_3=0$ is chosen with two votes while
other candidates get zero vote.
The choice $w_3=0$ gives
\begin{eqnarray*}
f_0^{(2)}&=&f_0^{(3)},\\
f_1^{(2)}&=&f_1^{(3)},\\
g_0^{(2)}&=&g_0^{(3)},\\
g_1^{(2)}&=&g_1^{(3)}.
\end{eqnarray*}

At the iteration for $s=2$ the second
temination criterion is satisfied.
The computation process and the
decoded codeword by the second criterion at $s=2$
are the same as those by first criterion at $s=3$.

In the voting for $s=2$, $w_2=0$ is chosen with two votes while
other candidates get zero vote.
The choice $w_2=0$ gives
\begin{eqnarray*}
f_0^{(1)}&=&f_0^{(2)},\\
f_1^{(1)}&=&f_1^{(2)},\\
g_0^{(1)}&=&g_0^{(2)},\\
g_1^{(1)}&=&g_1^{(2)}.
\end{eqnarray*}

In the voting for $s=1$, $w_1=1$ is chosen with two votes while
other candidates get zero vote.
The choice $w_1=0$ gives
\begin{eqnarray*}
f_0^{(0)}&=&uz+u,\\
f_1^{(0)}&=&uvz+uv,\\
g_0^{(0)}&=&z+u^3,\\
g_1^{(0)}&=&vz+u^3v.
\end{eqnarray*}

In the voting for $s=0$, $w_0=1$ is chosen with four votes while
other candidates get zero vote.
The choice $w_0=1$ gives
\begin{eqnarray*}
f_0^{(-1)}&=&uz,\\
f_1^{(-1)}&=&uvz,\\
g_0^{(-1)}&=&z+u^3+1,\\
g_1^{(-1)}&=&vz+u^3v+v.
\end{eqnarray*}

At $s=-1$, we find $\alpha_0=0$, $\alpha_1=u$
and $2 = -v_Q(\alpha_1) \leq \tau = 2$.
Thus the algorithm outputs $(w_s)_{s\in \Gamma}$ as decoded information
by the third termination criterion.
The decoded codeword is again $(1$, $1$, $1$, $1$, $0$, $0$, $0$, $0)$.
The three termination criterion give the same decoded codeword,
but their numbers of iterations are different.

Recall that all of candidates $w_4=0$, $1$, $\beta$, $\beta^2$
are chosen by voting at $s=4$
because all the candidates have the same number of votes.
Only $w_4=0$ gives the correct codeword, while
$w_4=\beta$ and $w_4=\beta^2$  give two other different incorrect codewords
whose Hamming distances are both $2$ from $\vec{r}$.
In this example, multiple candidates are chosen only at $s=4$.

\section{Theoretical Analysis of the Proposed Modification}\label{sec4}
In this section we prove that our modified algorithm
can find all the codewords within Hamming distance $\tau$
 from the received word $\vec{r}$.
We also give upper bounds on the number of iterations
in Section \ref{sec:nriteration}.

\subsection{Supporting Lemmas}\label{sec:sl}
In Section \ref{sec:sl} we shall introduce several lemmas
necessary in Sections \ref{sec:nrvotes}--\ref{sec:firstwhy}.
Recall that the execution of our modified algorithm
can branch when there are multiple candidates satisfying
the condition (\ref{eq:acceptedvote}).
For a fixed sequence of determined $w_s$,
define $\vec{r}^{(N)} = \vec{r}$ and
recursively define $\vec{r}^{(\mathrm{prec}(s))} =
\vec{r}^{(s)} - \ev (w_s \varphi_s)$.
By definition $\vec{r}^{(-1)} = \vec{r}-\ev(\sum_{s\in \Gamma} w_s \varphi_s)$.

The following lemma explains why the authors include ``Gr\"obner bases''
in the paper title.
The module $I_{\vec{r}^{(N)}}$ was used in
\citet{beelen10,fujisawa06,lax12,lee09,mrg13jsc,sakata01,sakata03}
but the use of $I_{\vec{r}^{(s)}}$ with $s < N$ was new
in \citet{lee11}.
\begin{lem}\label{lem:ir}
Fix $s \in H(Q) \cup \{-1\}$.
Let $\vec{r}^{(s)}$ correspond to $w_s$ ($s \in \Gamma$) chosen
by the decoding algorithm.
Define the $\mathbf{F}_q[x_1]$-submodule $I_{\vec{r}^{(s)}}$ of
$\RR z \oplus \RR$ by
\begin{equation}
I_{\vec{r}^{(s)}}=\{ \alpha_0 + \alpha_1 z \mid \alpha_0, \alpha_1 \in \RR,
v_{P_i}(\alpha_0+r_i^{(s)}\alpha_1) \geq 1, 1 \leq i \leq n \},\label{eq:ir}
\end{equation}
where $\vec{r}^{(s)} = (r_1^{(s)}$, \ldots, $r_n^{(s)})$.
Then $\{ f_i^{(s)}$, $g_i^{(s)} \mid i=0$, \ldots, $a_1-1\}$
is a Gr\"obner basis of $I_{\vec{r}^{(s)}}$
with respect to $<_s$ as an $\mathbf{F}_q[x_1]$-module.
\end{lem}
\proofstart
This lemma is a generalization of \citet[Proposition 11]{lee11}.
We can prove this lemma in
exactly the same way as the proof of \citet[Proposition 11]{lee11}
with replacing $y^j$ in \citet{lee11} with $y_j$ and $s-1$ in \citet{lee11}
by $\mathrm{prec}(s)$.
\proofend

The following proposition shows that the original decoding algorithm
\citet{lee11} can correct errors up to half the bound
$d_{\mathrm{AG}}(C_\Gamma)$, which was not claimed in \citet{lee11}.
\begin{prop}\label{prop:AG}
Fix $s \in \Gamma$.
Let $\lambda(s)$ as defined in Eq.\ (\ref{eq:lambda}) and
$\nu(s)$ as defined in Eq.\ (\ref{eq:nu}).
Then $\nu(s) = \lambda(s)$.
\end{prop}
\proofstart
Let $T_i = \{ j \in H(Q) \mid j \equiv i \pmod{a_1}, j+s \in \widehat{H}(Q)\}$, then we have $\lambda(s) = \sharp T_0 + \cdots + \sharp T_{a_1-1}$.
Moreover, 
observe that
\[
H(Q) \setminus \widehat{H}(Q) =
\{ -v_Q (\eta_i x_1^k) \mid
i=0, \ldots, a_1-1, k=0,1,\ldots \}.
\]
Therefore, for $s \in \Gamma$ we have
\begin{eqnarray*}
T_i &=& \{ j \in H(Q) \mid j \equiv i \pmod{a_1},
j+s \in \widehat{H}(Q) \} \\
&=& \{ j \in H(Q) \mid j \equiv i \pmod{a_1},
j+s \notin H(Q) \setminus \widehat{H}(Q) \}\\
&=& \{ j \in H(Q) \mid j \equiv i \pmod{a_1},
j+s \notin \{ -v_Q(\eta_{i'} x_1^k) \mid
 k \ge 0 \} \}\\
&=& \{ -v_Q(y_i x_1^m) \mid s  -v_Q(y_i x_1^m) \notin \{ -v_Q(\eta_{i'} x_1^k) \mid  k \ge 0 \} \},
\end{eqnarray*}
where the third equality holds by  Eq.\ (\ref{eq:iprime}).
By the equalities above, we see
\[
\sharp T_i = \max\left\{0, \frac{-v_Q(\eta_{i'}) + v_Q (y_i)-s}{-v_Q(x_1)}\right\},
\]
which proves the equality $\nu(s)=\lambda(s)$.
\proofend

\citet{lee11} showed that their original decoding
algorithm can correct up to $\lfloor (d_{\mathrm{LBAO}}(C_u)-1)/2 \rfloor$
errors,
where $d_{\mathrm{LBAO}}(C_u)
=\min\{\nu(s)\mid s\in
H(Q)$, $s\le u\}$.
Proposition \ref{prop:AG} implies that $d_{\mathrm{LBAO}}(C_u)$
is equivalent to $d_{\mathrm{AG}} (C_u)$
for every one-point primal code $C_u$,
and therefore \citet[Theorem 8]{andersen08} implies \citet[Proposition 12]{lee11}.
In another recent paper \cite{gmr13}
we proved that $d_{\mathrm{AG}}$ and $d_{\mathrm{LBAO}}$
are equal to the Feng-Rao bound as defined
in \citet{beelen08,ldecodepaper} for $C_u$.

\subsection{Lower Bound for the Number of Votes}\label{sec:nrvotes}
In Section \ref{sec:nrvotes}
we discuss the number of votes (\ref{eq:acceptedvote})
which a candidate $w_{s,i}$ receives.
Since we study list decoding,
we cannot assume the original transmitted codeword nor
the error vector as in \citet{lee11}.
Nevertheless,
the original theorems in \citet{lee11} allow natural generalizations
to the list decoding context.

\begin{lem}\label{lem:nrvotes}
Fix $s \in \Gamma$.
For $s' \in \Gamma^{(> s)}$, fix a sequence of $w_{s'}$ chosen by
the decoding algorithm, and define $\vec{r}^{(s)}$ corresponding
to the chosen sequence of $w_{s'}$.
Fix $\omega_s \in \mathbf{F}_q$.
Let $\vec{e} = (e_1$, \ldots, $e_n)^T$
be a nonzero vector with the minimum Hamming weight
in the coset $\vec{r}^{(s)} -\ev(\omega_s\varphi_s)+ C_{s-1}$, where $C_{s-1}$ is as defined
in Eq.\ (\ref{eq:cu}).
Define
\begin{eqnarray*}
J_{\vec{e}} &=& \bigcap_{e_i\neq 0} \mathcal{L}(-P_i+\infty Q)\\
&=& \mathcal{L}\left(\infty Q - \sum_{e_i\neq 0} P_i\right) \mbox{ (by \citet{matsumoto99ldpaper})}.
\end{eqnarray*}
Let $\{\epsilon_0$, \ldots, $\epsilon_{a_1-1}\}$ be a Gr\"obner basis
for $J_{\vec{e}}$ as an $\mathbf{F}_q[x_1]$-module
with respect to $<_s$ (for any integer $s$),
such that $\textsc{lm}(\epsilon_j) = x_1^{k_j} y_j$.

Under the above notations,
we have
\begin{eqnarray*}
-v_Q(\epsilon_i)+v_Q(a_{i,i} y_i )&\geq&a_1\bar{c}_i,\\
\min\set{-v_Q(\epsilon_i)+s,-v_Q(\eta_{i'})}&\geq&-v_Q(d_{i',i'} y_{i'}),
\end{eqnarray*}
for  $i$ with $w_{s,i}\neq \omega_s$, and
\[
\min\{-v_Q(\epsilon_i)+s, -v_Q(\eta_{i'})\} \geq
-v_Q(d_{i',i'}y_{i'})-a_1 \bar{c}_i,
\]
for $i$ with $w_{s,i}= \omega_s$.
\end{lem}
\proofstart
The proof is the same as those of \citet[Propositions 7 and 8]{lee11},
with replacing $y^j$ in \citet{lee11} by $y_j$,
$\delta(\cdot)$ in \citet{lee11} by $-v_Q(\cdot)$.
\proofend

The following lemma is a modification to \citet[Proposition 9]{lee11}
for the list decoding.
\begin{lem}
We retain notations from Lemma \ref{lem:nrvotes}.
We have 
\begin{eqnarray*}
\lefteqn{a_1 \sum_{w_{s,i}=\omega_s}\bar{c}_i\geq  a_1 \sum_{w_{s,i}\neq\omega_s}\bar{c}_i
- 2a_1\wt(\vec{e})} \\
&& \mbox{ } + \sum_{0\le i<a_1}\max\set{-v_Q(\eta_{i'})+v_Q( y_i)-s,-v_Q(\epsilon_i)+v_Q( y_i )}.
\end{eqnarray*}
\end{lem}
\proofstart
Lemma \ref{lem:nrvotes} implies
\begin{eqnarray*}
	\sum_{w_{s,i}=\omega_s}a_1\bar{c}_i
	&\geq&\sum_{w_{s,i}=\omega_s}-v_Q(d_{i',i'} y_{i'} )-\min\set{-v_Q(\epsilon_i)+s,-v_Q(\eta_{i'})}\\
	&\geq&\sum_{0\le i<a_1}-v_Q(d_{i',i'} y_{i'} )-\min\set{-v_Q(\epsilon_i)+s,-v_Q(\eta_{i'})}
\end{eqnarray*}
and
\begin{eqnarray*}
\sum_{w_{s,i}\neq\omega_s}a_1\bar{c}_i&\le&\sum_{w_{s,i}\neq\omega_s}-v_Q(\epsilon_i)+v_Q(a_{i,i} y_i )\\
	&\le&\sum_{0\le i<a_1}-v_Q(\epsilon_i)+v_Q(a_{i,i} y_i ).
\end{eqnarray*}
Now we have a chain of inequalities
\begin{eqnarray}
&& \sum_{w_{s,i}=\omega_s}a_1\bar{c}_i - \sum_{w_{s,i}\neq\omega_s}a_1\bar{c}_i\nonumber\\
& \geq & 
\sum_{0\le i<a_1}-v_Q(d_{i',i'} y_{i'})-\min\set{-v_Q(\epsilon_i)+s,-v_Q(\eta_{i'})}\nonumber\\
	&&- \sum_{0\le i<a_1}-v_Q(\epsilon_i)+v_Q(a_{i,i} y_i )\nonumber\\
&=& \sum_{0\le i<a_1}-v_Q(d_{i',i'} y_{i'} ) -v_Q(a_{i,i} y_i)\nonumber
\\&&\mbox{ }-\min\set{-v_Q(\epsilon_i)+s,-v_Q(\eta_{i'})}+v_Q(\epsilon_i)\nonumber\\
&=	& \sum_{0\le i<a_1}-v_Q(\eta_{i'})-v_Q( y_i )\label{eq:this}\\
&&\mbox{ }+\max\set{+v_Q(\epsilon_i)-s,+v_Q(\eta_{i'})}+v_Q(\epsilon_i)\nonumber\\
&=	& \sum_{0\le i<a_1}\max\set{-v_Q(\eta_{i'})+v_Q( y_i )-s,-v_Q(\epsilon_i)+v_Q( y_i )}\nonumber\\
&&	-\sum_{0\le i<a_1}2(-v_Q(\epsilon_i)+v_Q( y_i ))\nonumber
\end{eqnarray}  
where at Eq.\ (\ref{eq:this}) we used the equality
\begin{eqnarray*}
&& 	\sum_{0\le i<a_1}-v_Q(d_{i',i'} y_{i'} )+\sum_{0\le i<a_1}-v_Q(a_{i,i} y_i)\\
&=&\sum_{0\le i<a_1}-v_Q(d_{i,i} y_i )+\sum_{0\le i<a_1}-v_Q(a_{i,i} y_i )\\
&=&\sum_{0\le i<a_1}(-v_Q(d_{i,i})-v_Q(a_{i,i}))+\sum_{0\le i<a_1}-2v_Q( y_i)\\
	&=&a_1n+\sum_{0\le i<a_1}-2v_Q( y_i )\\
&=&\sum_{0\le i<a_1}(-v_Q(\eta_{i})+v_Q(y_i))+\sum_{0\le i<a_1}-2v_Q(y_i)\\
&=&\sum_{0\le i<a_1}-v_Q(\eta_{i'})+\sum_{0\le i<a_1}-v_Q(y_i)
\end{eqnarray*}
shown in \citet[Lemma 2 and Eq.\ (1)]{lee11}. Finally note that
\begin{eqnarray*}
&&\sum_{0\le i<a_1}2(-v_Q(\epsilon_i)+v_Q(y_i))\\
&=&\sum_{0\le i<a_1}2a_1\deg_{x_1}(\epsilon_i)=2a_1\wt(\vec{e})
\end{eqnarray*}
by \citet[Eq.\ (3)]{lee11}.
\proofend

The following lemma is a modification to \citet[Proposition 10]{lee11}
for list decoding, and provides a lower bound
for the number of votes (\ref{eq:acceptedvote})
received by any candidate $\omega_s \in \mathbf{F}_q$,
as indicated in the section title.
\begin{prop}\label{prop:nrvotes}
We retain notations from Lemma \ref{lem:nrvotes}.
Let $\nu(s)$ be as defined in Eq.\ (\ref{eq:nu}).
We have
\[
	\sum_{w_{s,i}=\omega_s}\bar{c}_i\geq \sum_{w_{s,i}\neq\omega_s}\bar{c}_i - 2\wt(\vec{e})+\nu(s).
\]
\end{prop}
\proofstart
We have
\begin{eqnarray*}
&&\sum_{0\le i<a_1}\max\set{-v_Q(\eta_{i'})+v_Q(y_i)-s,-v_Q(\epsilon_i)+v_Q(y_i)}\\
&\ge&\sum_{0\le i<a_1}\max\set{-v_Q(\eta_{i'})+v_Q(y_i)-s,0}
\end{eqnarray*}
as $-v_Q(\epsilon_i)+v_Q(y_i)\ge 0$ for $0\le i<a_1$.
\proofend

\subsection{Correctness of the Modified List Decoding Algorithm with
the Third Iteration Termination Criterion}\label{sec:thirdwhy}
In this subsection and the following sections,
we shall prove that the proposed list decoding algorithm
will find all the codewords within the Hamming distance $\tau$
 from the received word $\vec{r}$.
Since the third iteration termination criterion is the
easiest to analyze,
we start with the third one.

Fix a sequence $w_s$ for $s\in \Gamma$.
If $\wt(\vec{r} - \ev(\sum_{s\in \Gamma} w_s \varphi_s))\leq \tau$
then the sequence $w_s$ is found by the algorithm
because of 
Proposition \ref{prop:nrvotes}.
When $2 \tau < d_{\mathrm{AG}} (C_\Gamma )$,
by Proposition \ref{prop:AG} the decoding is not list decoding,
and the algorithm just declares the sequence $w_s$
as the transmitted information.

On the other hand, if
$2 \tau \geq d_{\mathrm{AG}} (C_\Gamma )$,
then the found sequence could correspond to a codeword
more distant than Hamming distance $\tau$,
and the algorithm examines the Hamming distance
between the found codeword and the received word $\vec{r}$.

Since computing $\ev(f)$ for $f\in \RR$ needs many multiplications
in $\mathbf{F}_q$, the algorithm checks some sufficient conditions
to decide the Hamming distance
between the found codeword and the received word $\vec{r}$.
Let $\vec{r}^{(-1)} = (r^{(-1)}_1$, \ldots, $r^{(-1)}_n)$.
When $\alpha_0=0$ in Section \ref{sec:3rd},
by Lemma \ref{lem:ir}, we have
\begin{eqnarray*}
\wt(\vec{r} - \ev(\sum_{s\in \Gamma} w_s \varphi_s))&=& \wt(\vec{r}^{(-1)})\\
&\leq& \sum_{r^{(-1)}_i \neq 0} v_{P_i}(\alpha_1)\\
&\leq&-v_Q(\alpha_1),
\end{eqnarray*}
because Eq.\ (\ref{eq:ir}) and $\alpha_0=0$ implies
that $v_{P_i}(\alpha_1) \geq 1$ for $r^{(-1)}_i \neq 0$.
By the above equation,
$-v_Q(\alpha_1) \leq \tau$ implies that the found codeword
is within Hamming distance $\tau$ from $\vec{r}$.
This explains why the algorithm can avoid computation of the
evaluation map $\ev$ in Step \ref{third1} in Section \ref{sec:3rd}.

In order to explain Step \ref{third2}  in Section \ref{sec:3rd},
we shall show that the condition of Step \ref{third2}  in Section \ref{sec:3rd}
implies that
$\wt(\vec{r} - \ev(\sum_{s\in \Gamma} w_s \varphi_s))> \tau$.
Suppose that
$\wt(\vec{r} - \ev(\sum_{s\in \Gamma} w_s \varphi_s)) \leq \tau$.
Then there exists $\beta_1 \in \RR$ such that
$v_{P_i}(\beta) \geq 1$ for $r^{(-1)}_i \neq 0$,
$-v_Q(\beta) \leq \tau + g$, and $\beta_1 z \in I_{\vec{r}^{(-1)}}$.
Because the leading term of $\beta_1 z$
must be divisible by $\textsc{lt}(f_i^{(-1)})$ for some $i$
by the property of Gr\"obner bases,
we must have $-v_Q(\alpha_1) \leq -v_Q(\beta_1)$.
This explains why the algorithm can avoid computation of the
evaluation map $\ev$ in Step \ref{third2} in Section \ref{sec:3rd}.

Otherwise, the algorithm computes the Hamming distance between
the found codeword and $\vec{r}$ in Steps \ref{third3}
and \ref{third4} in Section \ref{sec:3rd}.

\subsection{Correctness of the Modified List Decoding Algorithm with
the Second Iteration Termination Criterion}\label{sec:secondwhy}
We shall explain why the second criterion in Section \ref{sec:second}
correctly finds the required codewords.
For explanation, we present slightly rephrased version of
facts in \citet{beelen08}.
\begin{lem}\label{lem:23}\citep[Lemma 2.3]{beelen08}
Let $\beta_1 z + \beta_0 \in I_{\vec{r}^{(s)}}$
with $\textsc{lt}(\beta_1 z + \beta_0)
= \textsc{lt}(\beta_1 z)$ with respect to $<_s$ and 
$-v_Q(\beta_1) < n-\tau-s$.
If there exists $f \in \mathcal{L}(sQ)$ such that
$\wt(\ev(f)-\vec{r}^{(s)}) \leq \tau$,
then we have $f = - \beta_0/\beta_1$.
\end{lem}
\proofstart
Observe that
$\textsc{lt}(\beta_1 z + \beta_0)
= \textsc{lt}(\beta_1 z)$
implies that $-v_Q(\beta_0) \leq -v_Q(\beta_1)+s
< n-\tau$.
The claim of Lemma \ref{lem:23} is equivalent to
\citet[Lemma 2.3]{beelen08} with $A = (n-\tau-1)Q$
and $G = sQ$.
Note that the assumption $\deg A > (n+\deg G)/2+g-1$ was not used
in \citet[Lemma 2.3]{beelen08} but
only in \citet[Lemma 2.4]{beelen08}.
\proofend
Note that the following proposition was essentially proved
in \citet[Proposition 2.10]{beelen08},
\citet[Section 14.2]{justesen04},
and \citet[Theorem 2.1]{shokrollahi99} with $b=1$.
\begin{prop}\label{prop:candivide}
Let $\alpha_0$ and $\alpha_1$ be as in Section \ref{sec:second}.
If $s < n -g -2\tau$ and there exists $f \in \mathcal{L}(sQ)$
such that $\wt(\ev(f)-\vec{r}^{(s)}) \leq \tau$,
then we have $f = -\alpha_0/\alpha_1$.
\end{prop}
\proofstart
Let $g \in \RR$ such that
$g(P_i) = 0$ if $f(P_i) \neq r_i^{(s)}$,
and assume that $g$ has the minimum
pole order at $Q$ among such elements in $\RR$.
Then $-v_Q(g) \leq \tau + g$. One has that
$gz - fg \in I_{\vec{r}^{(s)}}$ and
$\textsc{lt}(gz - fg) = \textsc{lt}(gz)$ with respect to $<_s$.
By the property of Gr\"obner bases,
$\textsc{lt}(gz)$ is divisible by $\textsc{lt}(f^{(s)}_i)$
for some $i$, which implies $-v_Q(\alpha_1) \leq 
-v_Q(g) \leq \tau + g$.
By Lemma \ref{lem:23} we have $f = -\alpha_0/\alpha_1$.
\proofend

We explain how the procedure in Section \ref{sec:second}
works as desired.
When the condition in Step \ref{second1} in Section \ref{sec:second}
is true, then there cannot be a
codeword within Hamming distance
$\tau$ from $\vec{r}^{(s)}$ by the same reason as Section \ref{sec:thirdwhy}.
So the algorithm stops processing with $\vec{r}^{(s)}$.

When $2 \tau < d_{\mathrm{AG}} (C_\Gamma )$,
then the algorithm declares 
$-\alpha_0/\alpha_1+\sum_{s' \in \Gamma^{(> s)}} w_{s'} \varphi_{s'}$
as the unique codeword.

When $2 \tau \geq d_{\mathrm{AG}} (C_\Gamma )$,
then the algorithm examines the found codeword close enough to
$\vec{r}$ in Steps \ref{second11} and \ref{second12} in Section \ref{sec:second}.
When $-v_Q(\alpha_1) \leq \tau$
we can avoid computation of the evaluation map $\ev$ by the
same reason as Section \ref{sec:thirdwhy},
which is checked at Step \ref{second11}.
Otherwise we compute the codeword vector at Step \ref{second12}
and examine its Hamming distance to $\vec{r}^{(s)}$.

By Proposition \ref{prop:candivide},
the codeword must be found at $s= \max\{ s' \in \Gamma 
\mid s' < n-2 \tau + g\}$. Therefore, we do not execute the
iteration at $s< \max\{ s' \in \Gamma 
\mid s'< n-2 \tau + g \}$.

\subsection{Correctness of the Modified List Decoding Algorithm with
the First Iteration Termination Criterion}\label{sec:firstwhy}
We shall explain why the first criterion in Section \ref{sec:first}
correctly finds the required codewords.
The idea behind the first criterion is that
there cannot be another codeword within Hamming distance $\tau$
 from $\vec{r}^{(s)}$ when the algorithm already found one.
So the algorithm can stop iteration with smaller $s$
once a codeword is found as
$\ev(-\alpha_0/\alpha_1 +\sum_{s' \in \Gamma^{(> s)}} w_{s'} \varphi_{s'})$.

The algorithm does not examine conditions when
$-v_Q(\alpha_1) > \tau + g$ by the same reason as
Sections \ref{sec:thirdwhy} and \ref{sec:secondwhy}.
When $2 \tau < d_{\mathrm{AG}} (C_\Gamma )$,
then the algorithm declares 
$-\alpha_0/\alpha_1+\sum_{s' \in \Gamma^{(> s)}} w_{s'} \varphi_{s'}$
as the unique codeword.

When $2 \tau \geq d_{\mathrm{AG}} (C_\Gamma )$,
then the algorithm examines the found codeword close enough to
$\vec{r}$ in Steps \ref{first11}--\ref{first13} in Section \ref{sec:first}.
When $-v_Q(\alpha_1) \leq \tau$
we can avoid computation of the evaluation map $\ev$ by the
same reason as Section \ref{sec:thirdwhy},
which is checked at Step \ref{first11}.

By Proposition \ref{prop:candivide},
the codeword must be found at some $s\geq \max\{ s' \in \Gamma 
\mid s' < n-2 \tau + g\}$. Therefore, we do not execute the
iteration at $s< \max\{ s' \in \Gamma 
\mid s' < n-2 \tau + g\}$.

\subsection{Upper Bound on the Number of Iterations and the Worst Case Complexity}\label{sec:nriteration}
Observe that for $s > \max \Gamma$ we set always $w_s$ to $0$.
For each $s \in \Gamma$ satisfying
$\nu(s) \leq 2 \tau$,
the number of accepted candidates satisfying
Eq.\ (\ref{eq:acceptedvote})
can be at most $q$.
On the other hand, for $s$ with $\nu(s) > 2\tau$,
the number of candidates is either zero or one,
because at most one $w \in \mathbf{F}_q$ can satisfy Eq.\ 
(\ref{eq:acceptedvote}).
Therefore, we have upper bounds for the number of iterations,
counting executions of Rebasing in Section \ref{sec:rebasing},
as
\begin{eqnarray}
&&\sharp \{s\in H(Q) \mid \max \Gamma \leq s < N\}
+ \exp_q (\sharp \{s \in \Gamma \mid \nu(s) \leq 2 \tau\})\nonumber\\
&&\times \sharp \{s \in H(Q)\cup\{-1\} \mid
s < \max \Gamma \}\label{iteration-ub1}
\end{eqnarray}
for the third criterion for judging termination,
where $\exp_q(x) = q^x$, and
\begin{eqnarray}
&\sharp \{s\in H(Q) \mid \max \Gamma \leq s < N)\}
+ \exp_q (\sharp \{s \in \Gamma \mid \nu(s) \leq 2 \tau\})\nonumber\\
&\times \sharp \{s \in H(Q) \mid
\max\{ s' \in \Gamma 
\mid  s' < n-2 \tau - g\} \leq s < \max \Gamma \}\label{iteration-ub2}
\end{eqnarray}
for the first and the second criteria for judging termination.
We will use $\max \widehat{H}(Q)$
in place of $N$ in Eqs.\ (\ref{iteration-ub1}) and (\ref{iteration-ub2})
for computation in Tables \ref{tab1}--\ref{tab3},
because $N$ depents on $\vec{r}$ and $N \leq \max \widehat{H}(Q)$.

The proposed algorithm processes $2a_1$ polynomials (elements in $\RR$)
in each iteration and each polynomial has $O(n)$ terms.
For a precise closed-form evaluation of Eqs.\ (\ref{iteration-ub1}) and (\ref{iteration-ub2}),
we need a closed-form  upper bound of
$\sharp \{s \in \Gamma \mid \nu(s) \leq 2 \tau\}$ in terms of $\tau$,
but the authors could not find such one.
Clearly $n$ is an upper bound on
$\sharp \{s \in \Gamma \mid \nu(s) \leq 2 \tau\}$,
and the number of iterations is $O(\exp_q(n))$.
Therefore the worst-case complexity is $O(a_1 n \exp_q(n))$.
Nonetheless, we will see that the actual computational cost
can be lower than the existing list decoding algorithms
in some cases in Section \ref{sec:experiment}.
We remark that when $\tau < d_{\mathrm{AG}}(C_\Gamma)$ the
number of iterations is $\leq n$ and the worst case complexity is
$O(a_1 n^2)$ because the number of chosen candidate is at most
one by voting at each iteration,
and that the complexity $O(a_1 n^2)$
  is the same as the BMS algorithm \citep{sakata95b,sakata95a}.

Observe that
the list decoding can be implemented as
$\exp_q (\sharp \{s \in \Gamma \mid \nu(s) \leq 2 \tau\})$ parallel
execution of the unique decoding.
Therefore,
when one can afford 
$\exp_q (\sharp \{s \in \Gamma \mid \nu(s) \leq 2 \tau\})$
parallel implementation,
which increases the circuit size,
the decoding time of list decoding is the same as that of the unique
decoding.

\begin{table*}[!t]
\caption{Decoding results of codes on the Klein quartic ($\mathbf{F}_q=\mathbf{F}_8$, $g=3$ and $n=23$)}\label{tab1}
\centering
\rotatebox{270}{%
\footnotesize
\begin{tabular}{|ccccrrrrrrr|}\hline
$\sharp \Gamma$&$d_{\mathrm{AG}}(C_\Gamma)$&\# Errors&Termination&\multicolumn{3}{c}{\# Iterations}&\multicolumn{2}{c}{\# Multiplications}&\multicolumn{2}{c|}{\# Codewords}\\
&&$=\tau$&Criterion&\multicolumn{3}{c}{}&\multicolumn{2}{c}{\& Divisions in $\mathbf{F}_q$}&\multicolumn{2}{c|}{Found}\\\cline{5-11}
&&&in Sec.\ \ref{sec:termination}&Eq.(\ref{iteration-ub1}),(\ref{iteration-ub2})&Avg.&Max.&Avg.&Max.&Avg.&Max.\\\hline
18&4&1&1st&11&8.00&8&1,170.09&1,254&1.00&1\\
&&&2nd&11&11.00&11&844.98&879&&\\
&&&3rd&26&26.00&26&976.32&1,018&&\\\cline{3-11}
&&2&1st&328
&196.63&260&26,203.96&77,209&1.34&3\\
&&&2nd&328&200.64&269&8,349.67&15,457&&\\
&&&3rd&1,160
&219.07&313&7,813.76&10,083&&\\\cline{3-11}
&&3&1st&28,680
&11,996.34&13,353&1,626,658.69&2,490,386&19.75&28\\
&&&2nd&28,680&12,055.56&13,419&608,535.03&711,315&&\\
&&&3rd&73,736
&12,436.00&13,853&580,504.03&642,419&&\\\hline
11&10&4&1st&17&14.64&15&1,324.76&1,484&1.00&1\\
&&&2nd&17&16.64&17&1,161.52&1,293&&\\
&&&3rd&26&25.64&26&1,329.07&1,468&&\\\cline{3-11}
&&5&1st&47
&35.20&44&3,673.78&5,549&1.00&1\\
&&&2nd&47&38.20&47&2,915.04&3,622&&\\
&&&3rd&103
&45.41&72&3,072.08&3,769&&\\\cline{3-11}
&&6&1st&3,087
&1,507.95&1,692&164,274.07&188,797&1.11&3\\
&&&2nd&3,087&1,511.28&1,695&113,592.10&130,810&&\\
&&&3rd&5,647
&1,535.23&1,725&113,472.30&130,697&&\\\hline
\end{tabular}}
\end{table*}

\begin{table*}[t!]
\caption{Decoding results of the one-point Hermitian codes ($\mathbf{F}_q=\mathbf{F}_{16}$, $g=6$, $n=64$ and $\sharp \Gamma=55$). The meanings of $N$ and $R$ in
the third column is explained in Section \ref{sec51}.}\label{tab2}
\centering
\rotatebox{270}{%
\footnotesize
\begin{tabular}{|ccccrrrrrrr|}\hline
$\sharp \Gamma$&$d_{\mathrm{AG}}(C_\Gamma)$&\# Errors&Termination&\multicolumn{3}{c}{\# Iterations}&\multicolumn{2}{c}{\# Multiplications}&\multicolumn{2}{c|}{\# Codewords}\\
&&$=\tau$&Criterion&\multicolumn{3}{c}{}&\multicolumn{2}{c}{\& Divisions in $\mathbf{F}_q$}&\multicolumn{2}{c|}{Found}\\\cline{5-11}
&&&in Sec.\ \ref{sec:termination}&Eq.(\ref{iteration-ub1}),(\ref{iteration-ub2})&Avg.&Max.&Avg.&Max.&Avg.&Max.\\\hline
55&6&2R&1st&22&16.38&17&9,049.77&9,495&1.00&1\\
&&&2nd&22&21.79&22&5,483.91&5,614&&\\
&&&3rd&70&69.79&70&6,331.16&6,530&&\\\cline{3-11}
&&2N&1st&22&16.22&17&9,005.92&9,477&1.00&1\\
&&&2nd&22&21.22&22&5,414.32&5,607&&\\
&&&3rd&70&69.22&70&6,291.44&6,527&&\\\cline{3-11}
&&3R&1st&2,829&796.17&1,139&289,992.45&2,154,489&1.00&2\\
&&&2nd&2,829&800.66&1,143&116,784.32&164,299&&\\
&&&3rd&14,605&846.78&1,191&117,848.30&130,366&&\\\cline{3-11}
&&3N&1st&2,829&750.73&817&334,588.68&360,413&2.28&5\\
&&&2nd&2,829&761.79&825&120,575.80&131,791&&\\
&&&3rd&14,605&872.97&917&119,940.46&134,297&&\\\cline{3-11}
&&4R&1st&851,981&21,376.57&33,187&8,012,813.23&14,988,534&1.48&4\\
&&&2nd&851,981&21,384.19&33,198&2,431,318.50&3,763,057&&\\
&&&3rd&3,735,565&21,458.60&33,327&2,432,782.60&3,761,206&&\\\cline{3-11}
&&4N&1st&851,981&21,744.53&32,943&10,952,709.73&16,938,498&4.29&5\\
&&&2nd&851,981&21,769.88&32,962&2,457,072.25&3,801,066&&\\
&&&3rd&3,735,565&21,985.53&33,174&2,439,145.90&3,805,702&&\\\hline
\end{tabular}}
\end{table*}

\begin{table*}[t!]
\caption{Decoding results of the one-point Hermitian codes ($\mathbf{F}_q=\mathbf{F}_{16}$, $g=6$, $n=64$ and $\sharp \Gamma=39$). The meanings of $N$ and $R$ in
the third column is explained in Section \ref{sec51}.}\label{tab22}
\centering
\rotatebox{270}{%
\footnotesize
\begin{tabular}{|ccccrrrrrrr|}\hline
$\sharp \Gamma$&$d_{\mathrm{AG}}(C_\Gamma)$&\# Errors&Termination&\multicolumn{3}{c}{\# Iterations}&\multicolumn{2}{c}{\# Multiplications}&\multicolumn{2}{c|}{\# Codewords}\\
&&$=\tau$&Criterion&\multicolumn{3}{c}{}&\multicolumn{2}{c}{\& Divisions in $\mathbf{F}_q$}&\multicolumn{2}{c|}{Found}\\\cline{5-11}
&&&in Sec.\ \ref{sec:termination}&Eq.(\ref{iteration-ub1}),(\ref{iteration-ub2})&Avg.&Max.&Avg.&Max.&Avg.&Max.\\\hline
39&20&9R&1st&36&30.77&31&10,726.51&11,175&1.00&1\\
&&&2nd&36&35.77&36&8,851.66&9,249&&\\
&&&3rd&70&69.77&70&10,781.91&11,504&&\\\cline{3-11}
&&9N&1st&36&30.73&32&9,747.36&12,008&1.00&1\\
&&&2nd&36&35.73&36&7,807.55&8,378&&\\
&&&3rd&70&69.73&70&9,607.34&10,645&&\\\cline{3-11}
&&10R&1st&143&74.99&93&37,643.43&47,179&1.00&1\\
&&&2nd&143&80.99&99&22,792.85&25,972&&\\
&&&3rd&655&112.99&131&25,023.33&28,398&&\\\cline{3-11}
&&10N&1st&143&77.64&126&46,759.83&177,947&2.00&2\\
&&&2nd&143&89.61&138&23,971.51&43,332&&\\
&&&3rd&655&153.73&244&28,063.19&36,453&&\\\cline{3-11}
&&11R&1st&36,895&12,112.08&12,859&7,480,228.09&7,911,646&1.00&1\\
&&&2nd&36,895&12,118.08&12,865&3,186,977.77&3,352,388&&\\
&&&3rd&159,775&12,148.10&12,895&3,189,262.07&3,354,588&&\\\cline{3-11}
&&11N&1st&36,895&10,417.34&12,285&6,123,703.49&7,582,687&2.01&6\\
&&&2nd&36,895&10,429.34&12,297&2,638,130.92&3,226,308&&\\
&&&3rd&159,775&10,491.11&12,357&2,641,014.19&3,230,078&&\\\hline
\end{tabular}}
\end{table*}

\begin{table*}[!t]
\caption{Decoding results of codes on the curve in Example \ref{ex:gs1} ($\mathbf{F}_q=\mathbf{F}_9$, $g=22$ and $n=77$).
We note that Eq.\ (\ref{iteration-ub1}) give the same value for $\tau=10$ and $\tau=11$.}\label{tab3}
\centering
\rotatebox{270}{%
\footnotesize
\begin{tabular}{|ccccrrrrrrr|}\hline
$\sharp \Gamma$&$d_{\mathrm{AG}}(C_\Gamma)$&\# Errors&Termination&\multicolumn{3}{c}{\# Iterations}&\multicolumn{2}{c}{\# Multiplications}&\multicolumn{2}{c|}{\# Codewords}\\
&&$=\tau$&Criterion&\multicolumn{3}{c}{}&\multicolumn{2}{c}{\& Divisions in $\mathbf{F}_q$}&\multicolumn{2}{c|}{Found}\\\cline{5-11}
&&&in Sec.\ \ref{sec:termination}&Eq.(\ref{iteration-ub1}),(\ref{iteration-ub2})&Avg.&Max.&Avg.&Max.&Avg.&Max.\\\hline
58&6&2&1st&60&38.85&42&39,473.98&62,479&1.00&1\\
&&&2nd&60&59.30&60&12,255.73&13,348&&\\
&&&3rd&89&88.30&89&13,710.71&14,886&&\\\cline{3-11}
&&3&1st&2,862&62.50&120&36,350.41&72,463&1.00&1\\
&&&2nd&2,862&79.62&134&21,754.75&30,030&&\\
&&&3rd&5,049&106.62&161&23,556.44&31,555&&\\\hline
52&10&4&1st&64&46.94&51&37,228.61&78,090&1.00&1\\
&&&2nd&64&63.25&64&15,212.23&16,708&&\\
&&&3rd&89&88.25&89&17,082.34&18,879&&\\\cline{3-11}
&&5&1st&196,866&48.96&163&24,776.23&88,893&1.00&1\\
&&&2nd&196,866&66.17&178&20,660.52&69,176&&\\
&&&3rd&347,769&89.17&201&22,591.78&71,264&&\\\hline
37&20&9&1st&73&57.28&60&24,998.86&48,611&1.00&1\\
&&&2nd&73&72.34&73&23,168.98&24,784&&\\
&&&3rd&89&88.34&89&25,655.70&27,675&&\\\cline{3-11}
&&10&1st&1,915&58.67&61&25,492.43&43,294&1.00&1\\
&&&2nd&1,915&74.39&75&27,355.54&29,452&&\\
&&&3rd&3,049&88.39&89&29,678.00&31,836&&\\\cline{3-11}
&&11&1st&2,077&225.59&253&167,152.76&191,250&1.00&1\\
&&&2nd&2,077&242.36&268&124,664.66&139,519&&\\
&&&3rd&3,049&254.36&280&126,693.96&141,989&&\\\hline
\end{tabular}}
\end{table*}
\afterpage{\clearpage}

\section{Comparison to Conventional Methods}\label{sec:experiment}
\subsection{Simulation Condition and Results}\label{sec51}
We have provided an upper bound on the number of multiplications and
divisions at each step of the proposed algorithm.
We simulated $1,000$ transmissions of codewords
with the one-point primal codes on Klein quartic over $\mathbf{F}_8$
with $n=23$ by using Examples \ref{ex1} and \ref{ex2},
the one-point Hermitian codes over $\mathbf{F}_{16}$ with
$n=64$,
and the one-point primal codes on the curve in Example
\ref{ex:gs1} over $\mathbf{F}_9$ with $n=77$.

The program is implemented on the Singular computer algebra system
\citep{singular313}.
The program used for this simulation is available from
\url{http://arxiv.org/src/1203.6127v5/anc}.

In the execution, we counted the number of iterations,
(executions of Rebasing in Section \ref{sec:rebasing}),
the sum of upper bounds on the number of multiplications and
divisions
given in Eqs.\ (\ref{eq:compl:ini}),
(\ref{eq:compl3}), (\ref{eq:compl4}), (\ref{eq:compl5}), (\ref{eq:compl6}),
(\ref{eq:compl7}), (\ref{eq:compl1}), (\ref{eq:multirebasing})
and (\ref{eq:compl2}), and the number of codewords found.
Note also that Eq.\ (\ref{eq:NFmulti3}) instead of Eq.\ (\ref{eq:NFmulti4})
is used.

The parameter $\tau$ is set to the same as the number of
generated errors in each simulation condition.
$N$ or $R$ in the number of errors in Tables \ref{tab2} and \ref{tab22}
indicates that
the error vector is generated toward another codeword 
nearest from the transmitted codeword or
completely randomly, respectively.
The distribution of codewords is uniform on $C_\Gamma$.
That of error vectors is uniform on the vectors of
Hamming weight $\tau$.

In the code construction,
we always try to use the Feng-Rao improved construction.
Specifically, for a given designed distance $\delta$,
we choose $\Gamma = \{ s \in \widehat{H}(Q) \mid
\lambda(s)=\nu(s) \geq \delta \}$, and construct $C_\Gamma$ of
Eq.\ (\ref{eq:cgamma}).
In the following, the designed distance is denoted by
$d_{\mathrm{AG}}(C_\Gamma)$.
It can be seen from Tables \ref{tab1}--\ref{tab3}
and the following subsections that
the computational complexity of the proposed algorithm tends to
explode when the number of errors exceeds the error-correcting
capability of the Guruswami-Sudan algorithm \citep{guruswami99}.

\subsection{Comparison among the Three Proposed Termination Criteria}\label{sec:comparethree}
In Section \ref{sec:termination} we proposed three criteria for
terminating iteration of the proposed algorithm.
 From Tables \ref{tab1}--\ref{tab3},
one can see the following.
The first criterion has the smallest number of iterations,
and the second is the second smallest.
On the other hand, the first criterion has
the largest number of multiplications and divisions.
The second and the third have the similar numbers.
Only the first criterion was proposed in \citet{gmr12isit}
and we see that the new criteria are better than the old one.

The reason is as follows:
The computation of quotient
$\alpha_0/\alpha_1$ at Step \ref{first1} in
Section \ref{sec:first} is costlier than
updating $f_i^{(s)}$ and $g_i^{(s)}$
in Section \ref{sec:rebasing} and the first criterion
computes $\alpha_0/\alpha_1$ many times, which cancels
the effect of decrease in the number of iterations.
On the other hand, the second criterion computes $\alpha_0/\alpha_1$
only once, so it has the smaller number of multiplications and
divisions than the first.

The second criterion is faster when $2 \tau < d_{\mathrm{AG}}(C_\Gamma)$,
while the third tends to be faster when
$2 \tau \geq d_{\mathrm{AG}}(C_\Gamma)$.
In addition to this,
the ratio of the number of iterations in the second criterion to
that of the third is smaller with $2 \tau < d_{\mathrm{AG}}(C_\Gamma)$
than with $2 \tau \geq d_{\mathrm{AG}}(C_\Gamma)$.
We speculate the reason behind them as follows:
When $2 \tau \geq d_{\mathrm{AG}}(C_\Gamma)$ and a wrong candidate
is chosen at Eq.\ (\ref{eq:acceptedvote}), after several iterations
of Sections \ref{sec:termination} and \ref{sec:mainiteration},
we often observe in our simulation that no candidate satisfies
Eq.\ (\ref{eq:acceptedvote})
and the iteration stops automatically. Under such situation,
the second criterion does not help much to decrease
the number of iterations nor the computational
complexity when a wrong candidate is chosen at Eq.\ (\ref{eq:acceptedvote}),
and there are many occasions at which a wrong candidate is chosen
at Eq.\ (\ref{eq:acceptedvote}) when $2 \tau \geq d_{\mathrm{AG}}(C_\Gamma)$.
On the other hand,
when $2 \tau < d_{\mathrm{AG}}(C_\Gamma)$,
the second criterion helps to determine the transmitted information
earlier than the third.

\subsection{Tightness of Upper Bounds (\ref{iteration-ub1})
and (\ref{iteration-ub2})}
In Table \ref{tab3},
we observe that the upper bounds (\ref{iteration-ub1})
and (\ref{iteration-ub2}) are much larger than the
actual number of iterations for $\tau=5$.
The disappearance of candidates satisfying Eq.\ (\ref{eq:acceptedvote})
in the last paragraph
may also explain the reason behind the large differences for
$\tau=5$.

On the other hand, we observe that
the upper bound (\ref{iteration-ub2}) is quite tight
for $\tau=5$ in Table \ref{tab1} and $\tau=10$N in Table \ref{tab22}.
This suggests that improvement of Eq.\ (\ref{iteration-ub2})
may need some additional assumption.

\subsection{Klein Quartic, $(d_{\mathrm{AG}}(C_\Gamma),\tau)=(4,1)$ or
$(10,4)$}
We can use \citet{beelen07,beelen08,duursma11,duursma10,ldecodepaper} to decode this set of
parameters. It is essentially the forward elimination 
in the Gaussian elimination,
and it takes roughly $n^3/3$ multiplications.
In this case $n^3/3= 4,055$.
The proposed algorithm has lower complexity than
\citet{beelen07,beelen08,duursma11,duursma10,ldecodepaper}.

\subsection{Klein Quartic, $(d_{\mathrm{AG}}(C_\Gamma),\tau)=(4,2)$ or $(4,3)$}
The code is $C_u$ with $u=20$, $\dim C_u = 18$.
There is no previously known algorithm that can handle this case.

\subsection{Klein Quartic, $(d_{\mathrm{AG}}(C_\Gamma),\tau)=(10,5)$}
The code is $C_u$ with $u=13$, $\dim C_u = 11$.
According to \citet[Figure 1]{beelen10},
we can use the original Guruswami-Sudan \citep{guruswami99} but
it seems that its faster variants cannot be used.
We need  multiplicity $7$ to correct $5$ errors.
We have to solve a system of $23(7+1)7/2 = 644$
linear equations. It takes $644^3/3 =89,029,994$
multiplications
in $\mathbf{F}_8$. The proposed algorithm is much faster.

\subsection{Klein Quartic, $(d_{\mathrm{AG}}(C_\Gamma),\tau)=(10,6)$}
The code is $C_u$ with $u=13$, $\dim C_u = 11$.
There is no previously known algorithm that can handle this case.

\subsection{Hermitian, $(d_{\mathrm{AG}}(C_\Gamma),\tau)=(6,2)$ or
$(20,9)$}
We can use the BMS algorithm \citep{sakata95b,sakata95a} for
this case.
The complexity of \citet{sakata95b,sakata95a}
is estimated as $O(a_1 n^2)$ and $a_1 n^2 = 24,576$.
The complexity of the proposed algorithm seems comparable to
\citet{sakata95b,sakata95a}.
However, we are not sure which one is faster.

\subsection{Hermitian, $(d_{\mathrm{AG}}(C_\Gamma),\tau)=(6,3)$ or $(6,4)$}
The code becomes the Feng-Rao improved code with
designed distance $6$.
Its dimension is $55$.
In order to have the same dimension by $C_u$ we have to
set $u=60$, whose AG bound \citep{andersen08} is $4$ and the Guruwsami-Sudan
can correct up to 2 errors.
The proposed algorithm finds all codewords in the improved
code with $3$ and $4$ errors.

\subsection{Hermitian, $(d_{\mathrm{AG}}(C_\Gamma),\tau)=(20,10)$}
The code is $C_u$ with $u=44$.
The required multiplicity is $11$, and the required designed list
size is $14$.
The fastest algorithm for the interpolation step seems
\citet{beelen10}.
\citet[Example 4]{beelen10} estimates
the complexity of their algorithm
as $O(\lambda^5 n^2 (\log \lambda n)^2 \log(\log \lambda n))$,
where $\lambda$ is the designed list size.
Ignoring the log factor and assuming the scaling factor one in
the big-$O$ notation, the number of multiplications and divisions
is $\lambda^5 n^2 = 2,202,927,104$. The proposed algorithm
needs much fewer number of multiplications and divisions
in $\mathbf{F}_{16}$.

\subsection{Hermitian, $(d_{\mathrm{AG}}(C_\Gamma),\tau)=(20,11)$}
The Guruwsami-Sudan algorithm \citep{guruswami99} can correct up to $10$ errors
and there seems no previously known algorithm that can handle
this case.

\subsection{Garcia-Stichtenoth (Example \ref{ex:gs1}),
$(d_{\mathrm{AG}}(C_\Gamma),\tau) = (6,2)$, $(10,4)$, or $(20,9)$}
We can use
\citet{beelen07,beelen08,duursma11,duursma10,ldecodepaper}
to decode this set of
parameters. It is essentially the forward elimination 
in the Gaussian elimination,
and it takes roughly $n^3/3$ multiplications.
In this case $n^3/3= 152,177$.
The proposed algorithm has the lower complexity than
\citet{beelen07,beelen08,duursma11,duursma10,ldecodepaper}.

\subsection{Garcia-Stichtenoth (Example \ref{ex:gs1}),
$(d_{\mathrm{AG}}(C_\Gamma),\tau) = (6,3)$}
This is a Feng-Rao improved code with dimension $58$.
In order to realize a code with the same dimension,
we have to set $u=79$ in $C_u$.
The Guruswami-Sudan algorithm \citep{guruswami99} can correct no error in this set
of parameters. There seems no previously known algorithm that can handle
this case.

\subsection{Garcia-Stichtenoth (Example \ref{ex:gs1}),
$(d_{\mathrm{AG}}(C_\Gamma),\tau) = (10,5)$}
This is a Feng-Rao improved code with dimension $52$.
In order to realize a code with the same dimension,
we have to set $u=73$ in $C_u$.
The Guruswami-Sudan algorithm \citep{guruswami99} can correct $2$ errors in this set
of parameters. There seems no previously known algorithm that can handle
this case.

\subsection{Garcia-Stichtenoth (Example \ref{ex:gs1}),
$(d_{\mathrm{AG}}(C_\Gamma),\tau) = (20,10)$}
This is an ordinary one-point AG code $C_u$ with $u=58$ and
dimension $37$.
The Guruswami-Sudan algorithm \citep{guruswami99} can correct $10$ errors
with the multiplicity $154$ and the designed list size $178$.
We have to solve a system of $77 \times (154+1)154/2 = 918,995$
linear equations. It takes $918,995^3/3 =258,712,963,551,308,291$
multiplications
in $\mathbf{F}_9$. The proposed algorithm is much faster.

\subsection{Garcia-Stichtenoth (Example \ref{ex:gs1}),
$(d_{\mathrm{AG}}(C_\Gamma),\tau) = (20,11)$}
The Guruswami-Sudan algorithm \citep{guruswami99} can correct up to $10$ errors
and there seems no previously known algorithm that can handle
this case.

\section{Conclusion}\label{sec6}
In this paper, we modified the unique decoding algorithm
for plane AG codes in \citet{lee11} so that it can support
one-point AG codes on \emph{any} curve, and
so that it can do the list decoding.
The error correction capability of the original \citep{lee11}
and our modified algorithms are also expressed
in terms of the minimum distance lower bound in \citet{andersen08}.

We also proposed procedures to compute products and
quotients in coordinate ring of affine algebraic curves,
and by using those procedures we demonstrated that
the modified decoding algorithm can be executed quickly.
Specifically, its computational complexity is theoretically
  the same as
the BMS algorithm \citep{sakata95b,sakata95a} for one-point
Hermitian codes.
It is also much faster than the standard list decoding
  algorithms \citep{beelen10,guruswami99} for many cases
  that are examined and reported in our computational experiments
  in which examined error-correcting codes have midium sizes.
It should be noted that
  as a list decoding algorithm the proposed method seems to have
  exponential worst-case computational complexity while the previous
  proposals \citep{beelen10,guruswami99} have polynomial ones,
  and that the proposed method is expected to be slower than
  the previous proposal for very large/special inputs.

The original decoding algorithm \citep{lee11}
allows parallel implementation
on circuits like the K\"otter architecture \citep{koetter98}.
Our modified algorithm retains this advantage.
Moreover, if one can afford large circuit size,
the proposed list decoding algorithm can be executed as
quickly as the unique decoding algorithm by parallel implementation
on a circuit.

\section*{Acknowledgment}
The authors deeply thank the editor and anonymous reviewers
for their careful reading that improved the
presentation.
This research was partially supported by 
 the MEXT Grant-in-Aid for Scientific Research (A) Nos.\ 23246071 and
26289116, the
 Villum Foundation through their VELUX Visiting Professor Programme
 2011--2012 and 2014, the Danish National Research Foundation and the National
 Science Foundation of China (Grant No.\ 11061130539) for the
 Danish-Chinese Center for Applications of Algebraic Geometry in
 Coding Theory and Cryptography, the Danish Council for              
 Independent Research, grant DFF-4002-00367,
and the Spanish MINECO grant No.\ MTM2012-36917-C03-03.
The computer experiments in this research was conducted on
Singular 3.1.3 \citep{singular313}.


\end{document}